\documentclass{aa}

\usepackage{natbib}
\bibpunct{(}{)}{;}{a}{}{,} 

\usepackage{lscape}

\usepackage{placeins}
\usepackage{graphicx}
\usepackage{amsmath}
\usepackage{txfonts}
\usepackage{color}
\usepackage[dvipsnames]{xcolor}
\usepackage{subcaption}

\newcommand{\mdisk}[0]{\ensuremath{M_{\rm disk}}}

\newcommand{\alp}[0]{\ensuremath{\alpha_{\rm visc}}}
\newcommand{\tvisc}[0]{\ensuremath{t_{\rm visc}}}
\newcommand{\rgas}[0]{\ensuremath{R_{\rm CO,\ 90\%}}}
\newcommand{\rc}[0]{\ensuremath{R_{\rm c}}}
\newcommand{\adisk}[0]{$\alpha-$disk}
\newcommand{\macc}[0]{\ensuremath{\dot{M}_{\rm acc}}}
\newcommand{\rinit}[0]{\ensuremath{R_{\rm init}}}
\newcommand{\minit}[0]{\ensuremath{M_{\rm init}}}
\newcommand{\mstar}[0]{\ensuremath{M_{*}}}
\newcommand{\msun}[0]{\ensuremath{\mathrm{M}_{\odot}}}

\newcommand{\krijtprep}[0]{Krijt et al., in prep. }

\begin{document} 

    \title{Observed sizes of planet-forming disks trace viscous spreading}

   \author{L. Trapman \inst{1}
          \and
          G. Rosotti \inst{1}
          \and
          A.D. Bosman \inst{1,2}
          \and
          M.R. Hogerheijde \inst{1,3}
          \and
          E.F. van Dishoeck \inst{1,4}
          }

   \institute{
            Leiden Observatory, Leiden University, Niels Bohrweg 2, NL-2333 CA Leiden, The Netherlands \\
            \email{trapman@strw.leidenuniv.nl}
        \and
            Department of Astronomy, University of Michigan, 1085 S. University Ave, Ann Arbor, MI 48109
        \and
            Anton Pannekoek Institute for Astronomy, University of Amsterdam, Science Park 904, 1090 GE Amsterdam, The Netherlands
        \and
            Max-Planck-institute f\"{u}r Extraterrestrische Physik, Giessenbachstra{\ss}e, D-85748 Garching, Germany
            }
   \date{Received xx; accepted yy}

  \abstract
   {The evolution of protoplanetary disks is dominated by the conservation of angular momentum, where the accretion of material onto the central star is fed by viscous expansion of the outer disk or by disk winds extracting angular momentum without changing the disk size. 
   Studying the time evolution of disk sizes allows us therefore to distinguish between viscous stresses or disk winds as the main mechanism of disk evolution. Observationally, estimates of the size of the gaseous disk are based on the extent of CO sub-millimeter rotational emission, which, during the evolution, is also affected by the changing physical and chemical conditions in the disk.}
   {We study how the gas outer radius measured from the extent of the CO emission changes with time in a viscously expanding disk and investigate to what degree this observable gas outer radius is a suitable tracer of viscous spreading and whether current observations are consistent with viscous evolution.} 
   {For a set of observationally informed initial conditions we calculate the viscously evolved density structure at several disk ages and use the thermochemical code \texttt{DALI} to compute synthetic emission maps, from which we measure gas outer radii in a similar fashion to observations.}
   {
   The gas outer radii (\rgas) measured from our models match the expectations of a viscously spreading disk:
   \rgas\ increases with time and for a given time \rgas\ is larger for a disk with a higher viscosity \alp. 
   However, in the extreme case where the disk mass is low ($\mdisk \leq 10^{-4}\ \msun$) and \alp\ is high ($\geq 10^{-2}$), \rgas\ will instead decrease with time as a result of CO photodissociation in the outer disk. 
   For most disk ages \rgas\ is up to$\sim12\times$ larger than the characteristic size \rc\ of the disk, and \rgas/\rc\ is largest for the most massive disk. 
   As a result of this difference, a simple conversion of \rgas\ to \alp\ will overestimate the true \alp\ of the disk by up to an order of magnitude.
   Based on our models, we find that most observed gas outer radii in Lupus can be explained using viscously evolving disks that start out small $(\rc(t=0) \simeq 10\ \mathrm{AU})$ and have a low viscosity $(\alp = 10^{-4} - 10^{-3})$.
   } 
   {Current observations are consistent with viscous evolution, but expanding the sample of observed gas disk sizes to star-forming regions both younger and older would better constrain the importance of viscous spreading during disk evolution.
   }
   
   \maketitle
%
\section{Introduction}
\label{sec: introduction}

Planetary systems form and grow in protoplanetary disks. These disks provide the raw materials, both gas and dust, to form the increasingly diverse population of exoplanets and planetary systems that has been observed (see, e.g., \citealt{Benz2014,Morton2016,Mordasini2018}). 
The formation of planets and the evolution of protoplanetary disks are closely related.
While planets are forming, the disk is evolving around them, affecting the availability of material and providing constantly changing physical conditions around the planets. 
In a protoplanetary accretion disk, material is transported through the disk and accreted onto the star. 
Exactly which physical process dominates the angular momentum transport and drives the accretion flow, a crucial part of disk evolution, is still debated.

It is commonly assumed that disks evolve under influence of an effective viscosity, where viscous stresses and turbulence transport angular moment to the outer disk (see, e.g. \citealt{LyndenBellPringle1974,ShakuraSunyaev1973}). As a consequence of the outward angular momentum transport, the bulk of the mass moves inward and is accreted onto the star. The physical processes than constitute this effective viscosity are still a matter of debate, with magnetorotional instability being the most accepted mechanism (see, e.g. \citealt{BalbusHawley1991,BalbusHawley1998}).  

An alternative hypothesis is that angular momentum can be removed by disk winds rather than being transported through the disk (see, e.g., 
\citealt{Turner2014} for a review). The presence of a vertical magnetic field 
in the disk can lead to the development of a magnetohydrodynamic (MHD) disk wind. These disk winds remove material from the surface of the disk and are thus able to provide some or all of the angular momentum removal required to fuel stellar accretion (see, e.g. \citealt{Ferreira2006,Bethune2017,ZhuStone2017}). 
Direct observational evidence of such disk winds focuses on the inner part of the disk but it is unclear whether winds dominate the transport of angular momentum and therefore how much winds affect the evolution of the disk
(see, e.g. \citealt{Pontoppidan2011,Bjerkeli2016,Tabone2017,deValon2020}).

Observationally these two scenarios make distinctly different predictions on how the sizes of protoplanetary disks evolve over time. In the viscous disk theory, conservation of angular momentum will ensure that some parts of the disk move outward. As a result, disk sizes should grow with time. 
If instead disk sizes do not grow with time, disk winds are likely driving disk evolution. 
To distinguish between viscous evolution and disk winds we need to define a \emph{disk size}, which has to be measured or inferred from observed emission, and examine how it changes as a function of disk age.

With the advent of the Atacama Large Millimeter/sub-Millimeter Array (ALMA) it has become possible to perform large surveys of protoplanetary disks at high angular resolution. This has resulted in a large number of disks for which the extent of the millimeter continuum emission, the \emph{dust disk size}, can be measured (see, e.g \citealt{barenfeld2017, Cox2017,Tazzari2017, ansdell2018, Cieza2018, Long2019}). 
However, this continuum emission is predominantly produced by millimeter-sized dust grains, which also undergo radial drift as a result of the drag force from the gas, an inward motion that complicates the picture. 
Moreover, radial drift and radially dependent grain growth lead to a dependence between the extent of the continuum emission and the wavelength of the observations, with emission at longer wavelengths being more compact (see e.g., \citealt{Tripathi2018}).

\cite{Rosotti2019} used a modeling framework to study the combined effect of radial drift and viscous spreading on the observed dust disk sizes. 
They determined that in order to measure viscous spreading, the dust disk size has to be defined as a high fraction ($\geq 95\%$) of the total continuum flux. To ensure that this dust disk size is well characterized, the dust continuum has to be resolved with high S/N.
They showed that existing surveys lack the sensitivity to detect viscous spreading.

To avoid having to disentangle the effects of radial drift from viscous spreading we can instead measure a \emph{gas disk size} from rotational line emission of molecules like CO and CN, which are commonly observed in protoplanetary disks. 
A often used definition for the gas disk size is the radius that encloses 90\% of the total CO $J = 2\,-\,1$ flux (\rgas; see, e.g. \citealt{ansdell2018}). This radius encloses most ($>98\%$) of the disk mass and it has been shown that \rgas\ is not affected by dust evolution \citep{Trapman2019a}.
The longer integration time required to detect this emission in the outer disk means that significant samples of measured gas disk sizes are only now becoming available (see, e.g. \citealt{barenfeld2017,ansdell2018}). Using a sample of measured gas disk sizes collated from literature, \cite{NajitaBergin2018} showed tentative evidence that older Class II sources have larger gas disk sizes that the younger Class I sources, consistent with expectations for viscous spreading. 
It should be noted however that the gas disk sizes in their sample were measured using a variety of different tracers and observational definitions of the gas disk size.

When searching for viscous spreading using measured gas disk sizes it is important to keep in mind that these gas disk sizes are an observed quantity, measured from molecular line emission.
As the disk evolves, densities and temperatures change, affecting the column densities and excitation levels of the gas tracers used to measure the disk size. 
How well the observed gas outer radius traces viscous expansion has not been investigated in much detail. 

Time-dependent chemistry also affects the gas tracers like CO that we use to measure gas disk sizes. At lower densities, found in the outer disk and at a few scale-heights above the midplane of the disk, CO is destroyed through photo-dissociation by UV radiation. \cite{Trapman2019a} showed that \rgas\ traces the point in the outer disk where CO becomes photo-dissociated. 
Deeper in the disk, around the midplane where the temperature is low, CO freezes out onto dust grains. Once frozen out, CO can be converted into other molecules like CO$_2$, CH$_4$ and CH$_3$OH (see, e.g. \citealt{Bosman2018b,Schwarz2018}). These molecules have higher binding energies than CO and therefore stay frozen out at temperatures where CO would normally desorb back into the gas phase. 
Through this process gas-phase CO can be more than an order of magnitude lower than the abundance of $10^{-4}$ with respect to molecular hydrogen, which is the expected abundance when most of the volatile carbon is contained in CO (see, e.g. \citealt{Lacy1994}).

In this work, we set up disk models with observationally informed initial conditions, let the surface density evolve viscously and use the thermochemical code \texttt{DALI} \citep{Bruderer2012,Bruderer2013} to study how the CO $J=2\,-\,1$ intensity profiles, and the gas disk sizes derived from them, change over time. We then compare with existing observations to see if the observations are consistent with viscous evolution.
This work is structured as follows: 
We introduce the setup and assumptions in our modeling in Section \ref{sec: Model setup}. 
In Section \ref{sec: results} we show how well observed gas outer radii trace viscous evolution both qualitatively and quantitatively. In Section \ref{sec: discussion} we compare our models to observations. 
We also study how chemical CO depletion through grain-surface chemistry will affect our results. We discuss whether external photo-evaporation could explain the small observed gas disk sizes, compare our results to disk evolution driven by magnetic disks winds and we discuss whether including episodic accretion would affect our results. We conclude in Section \ref{sec: conclusions} with that measured gas outer radii can be used to trace viscous spreading in disks.

\section{Model setup}
\label{sec: Model setup}

The gas disk size is commonly obtained from CO rotational line observations, for example by measuring the radius that enclose 90\% of the CO flux (e.g. \citealt{ansdell2018}) or by fitting a powerlaw to the observed visibilities (e.g. \citealt{barenfeld2017}). In this work we use the radius that encloses 90\% of the $^{12}$CO 2-1 flux (\rgas) as the definition of the observed gas disk size. 
\cite{Trapman2019a} showed that \rgas\ traces a fixed surface density in the outer disk, where CO becomes photo-dissociated. To see if \rgas\ is a suitable tracer of viscous evolution, we are therefore interested in how the extent of the $^{12}$CO intensity changes over time in a viscously evolving disk.

Our approach for setting up our models is the following.
First, we obtain a set of initial conditions that reproduce current observed stellar accretion rates, assuming that the disks feeding the stellar accretion have evolved viscously. Next we calculated the time evolution of the surface density. For each set of initial conditions, we analytically calculate the surface density profile $(\Sigma (R,t))$ at 10 different disk ages. For each of these time-steps we use the thermochemical code \texttt{DALI} \citep{Bruderer2012,Bruderer2013} to calculate the current temperature and chemical structure of the disk at that age and create synthetic emission maps of $^{12}$CO, from which we measure \rgas.

\subsection{Viscous evolution of the surface density}
\label{sec: viscous surface density}

Accretion disks in which the disk structure is shaped by viscosity are often described using a $\alpha-$disk formalism \citep{ShakuraSunyaev1973}, parameterizing the kinematic viscosity as $\nu = \alpha c_s H$, where $c_s$ is the sound speed and $H$ is the height above the midplane \citep{Pringle1981}. For an \adisk, the self-similar solution for the surface density $\Sigma$ is given by \citep{LyndenBellPringle1974,Hartmann1998}

\begin{equation}
\label{eq: surface density}
    \Sigma_{\rm gas}(R) = \frac{\left(2-\gamma\right)\mdisk(t)}{2\pi\rc(t)^2} \left( \frac{R}{\rc(t)} \right)^{-\gamma} \exp\left[ -\left(\frac{R}{\rc(t)}\right)^{2-\gamma}\right],
\end{equation}
where $\gamma$ is set by assuming that the viscosity varies radially as $\nu\propto R^{\gamma}$ and \mdisk\ and \rc\ are the disk mass and the characteristic radius, respectively. 

Following \cite{Hartmann1998}, the time evolution of \mdisk\ and \rc\ is described by
\begin{align}
\label{eq: mass time evolution}
\mdisk(t) &= \mdisk(t=0) \left(1 + \frac{t}{\tvisc} \right)^{-\frac{1}{[2(2-\gamma)]}}
          = \minit \left(1 + \frac{t}{\tvisc} \right)^{-\frac{1}{2}}\\
\rc(t)    &= \rc(t=0) \left(1 + \frac{t}{\tvisc} \right)^{\frac{1}{(2-\gamma)}}
          \ \ \ \ \ \ \ \ = \rinit \left(1 + \frac{t}{\tvisc} \right),
\end{align} 
where \tvisc\ is the viscous timescale and we have defined short hands for  the initial disk mass $\minit\equiv\mdisk(t=0)$, the initial characteristic size $\rinit\equiv\rc(t=0)$. For the second step and the rest of this work we have assumed $\gamma = 1$, since for a typical temperature profile it corresponds to the case of a constant \alp.

Combining equations \eqref{eq: surface density}, \eqref{eq: mass time evolution} and (3) we obtain the time evolution of the surface density profile
\begin{align}
\label{eq: surface density time evolution}
    \Sigma_{\rm gas}(t,R) &= \frac{\mdisk(t)}{2\pi\rc(t)^2} \left( \frac{R}{\rc(t)} \right)^{-1} \exp\left[ -\left(\frac{R}{\rc(t)}\right)\right]\\
    &= \frac{\minit}{2\pi\rinit^2} \left(1 + \frac{t}{\tvisc} \right)^{-\frac{3}{2}} \left( \frac{R}{\rinit \left(1 + \frac{t}{\tvisc} \right)} \right)^{-1} \\
    &\times \exp\left[ -\left(\frac{R}{\rinit \left(1 + \frac{t}{\tvisc} \right)}\right)\right]
\end{align}

\subsection{Initial conditions of the models}
\label{sec: initial condition}

\begin{figure}[htb]
    \centering
    \includegraphics[width=\columnwidth]{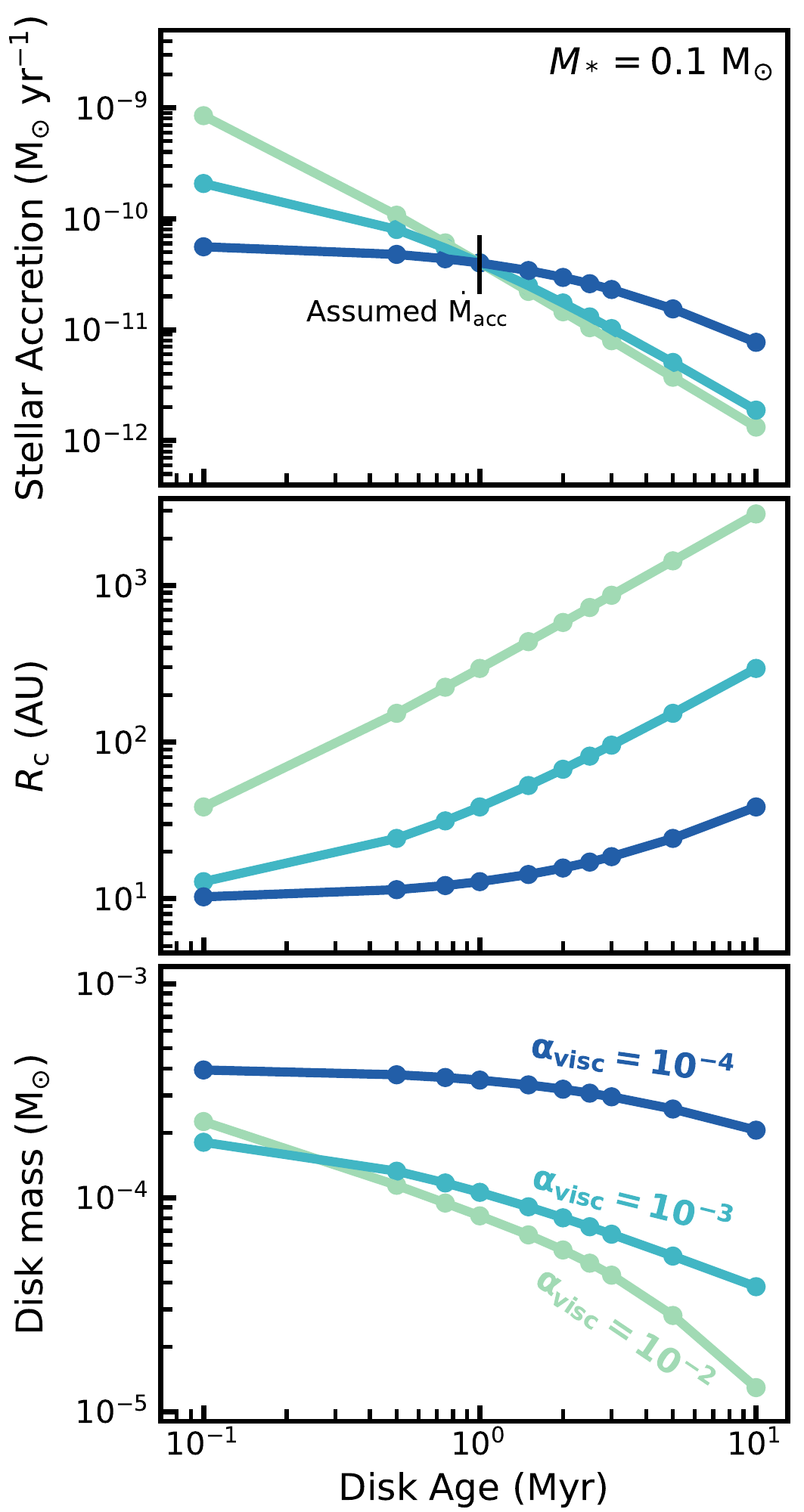}
    \caption{\label{fig: disk parameter evolution}
    Evolution of the disk parameters for the models with $M_* = 0.1\ \mathrm{M}_{\odot}$. Colours show different \alp. For reference, the viscous timescales are 0.046, 0.46 and 4.6 Myr for $\alp = 10^{-2}, 10^{-3}$ and $10^{-4}$, respectively. The evolution of the disk mass for the models with $\mstar = 0.32\ \msun$ and $1.0\ \msun$ is shown Figure \ref{fig: disk mass evolution}. Note that for the model with $\alp = 10^{-2}$, \mdisk\ starts out at higher mass compared to the model with $\alp = 10^{-3}$, but this model also loses mass at a faster rate.}
\end{figure}

For a viscous disk the initial disk mass \minit\ is related to the stellar accretion rate through \citep{Hartmann1998}
\begin{equation}
\label{eq: init disk - accretion}
    \minit = 2 \tvisc \macc(t) \left(\frac{t}{\tvisc} + 1 \right)^{3/2}.
\end{equation}
Under the assumption that the disk evolved viscously, we can calculated \minit\ given the stellar accretion rate measured at a time $t$ and the viscous timescale of the disk.
For various star forming regions, e.g. Lupus and Chamaeleon I, stellar accretion rates have been determined from observations (see, e.g \citealt{alcala2014,alcala2017,manara2017}). A correlation was found between the stellar mass \mstar\ and the stellar accretion rate \macc, best described by a broken powerlaw.
Based on equation \eqref{eq: init disk - accretion} a disk around a more massive star will therefore have a higher initial disk mass for the same viscous timescale. 

For our models we consider three stellar masses: $0.1, 0.32\ \mathrm{and}\ 1.0\ \mathrm{M}_{\odot}$. For each stellar mass, we use the observations, presented in Figure 6 of \cite{alcala2017}, to pick the
average stellar accretion rate associated to that stellar mass. 
For each stellar accretion rate, we calculate the initial disk mass using equation \eqref{eq: init disk - accretion} for 3 different viscous timescales.
The viscous timescale is computed for three values of the dimensionless viscosity $\alp = 10^{-2}, 10^{-3}, 10^{-4}$, assuming a characteristic radius of 10 AU (which is the radius we will employ; see below) and a disk temperature $T_{\rm disk}$ of 20 K (see, e.g. equation 37 in \citealt{Hartmann1998})
\begin{equation}
\frac{\tvisc}{\mathrm{yr}} = \frac{\rc^2}{\nu} \simeq 8\times10^{4} \left(\, \frac{\alp}{10^{-2}}\,\right)^{-1} \left(\,\frac{\rc}{10\ \mathrm{AU}}\,\right)\left(\,\frac{M_*}{0.5 \ \mathrm{M}_{\odot}} \,\right)^{1/2} \left(\,\frac{T_{\rm disk}}{10\ \mathrm{K}}\,\right)^{-1}.
\end{equation}

The combination of 3 stellar accretion rates and 3 viscous timescales results in 9 different disk models (see Table \ref{tab: model initial conditions}).
Figure \ref{fig: disk parameter evolution} shows how disk parameters like \macc, \rc\ and \mdisk\ evolve with time for the models with $\mstar = 0.1\ \msun$. Similarly, Figure \ref{fig: disk mass evolution} shows how \mdisk\ evolves for the models with $\mstar = 0.32\ \msun$ and $1.0\ \msun$. Note that the trends for \mdisk(t) are very similar, apart from starting at a higher initial \mdisk\ ($\sim$$10^{-2}\ \msun$ for $\mstar = 0.32\ \msun$ and $\sim$$10^{-1}\ \msun$ for $\mstar = 1.0\ \msun$; cf. Table \ref{tab: model initial conditions}).

\begin{figure*}[thb]
    \centering
    \includegraphics[width=\textwidth]{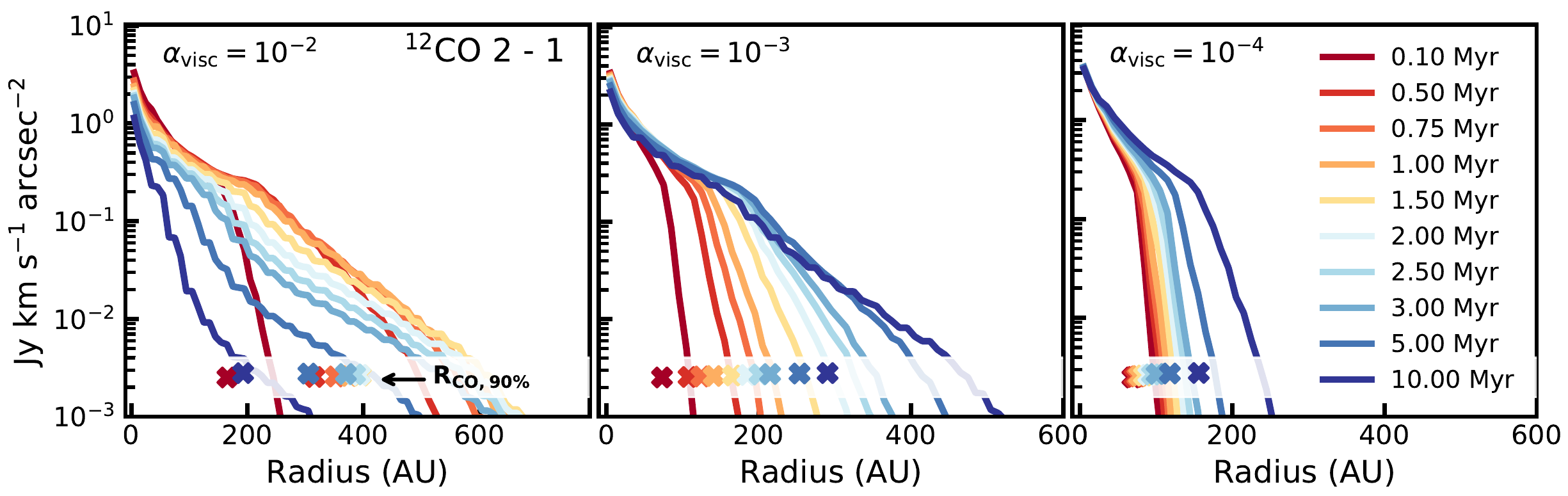}
    \caption{\label{fig: 12CO profile evolution}
    Evolution of the $^{12}$CO 2\,-\,1 radial intensity profiles for models with $M_* = 0.1\ \mathrm{M}_{\odot}$. Colours indicate different disk ages between 0.1 and 10 Myr. Crosses at the bottom of each panel show the gas outer radius, defined here as the radius that encloses 90\% of the total $^{12}$CO 2\,-\,1 flux. }
\end{figure*}

The initial size of disks is less well constrained, predominately due to a lack of high resolution observations for younger Class 1 and 0 objects. Recently, \citealt{Tobin2020} presented the VANDAM II survey: 330 protostars in Orion observed at 0.87 millimeter with ALMA at a resolution of $\sim0\farcs1$ ($\sim40$ AU in diameter). By fitting a 2D Gaussian to their dust millimeter observations, the authors determined median dust disk radii of $44.9^{+5.8}_{-3.4}$ AU and $37.0^{+4.9}_{-3.0}$ AU for their Class 0 and Class 1 protostars, suggesting that the majority of disks are initially very compact. 
It should be noted here that it is unclear whether the extent of the dust emission can be directly related to the gas disk size. However, there also similar evidence from the gas that Class 1 and 0 objects are compact. As part of the CALYPSO large program \cite{Maret2020} observed 16 Class 0 protostars and found that only two sources show Keplerian rotation at $\sim 50$ AU scales, suggesting that Keplerian disks larger than 50 AU, such as found for VLA 1623 \citep{Murillo2013}, are uncommon. 
We therefore adopt an initial disk size of $\rinit = 10\ \mathrm{AU}$ for our models. In Section \ref{sec: larger initial disk sizes} we discuss the impact of this choice. 

\begin{table}[htb]
    \centering
    \caption{\label{tab: model initial conditions} Initial conditions of our \texttt{DALI} models}
    \begin{tabular}{lc|ccc}
    \hline
    \hline
    $\mstar$& $(\mathrm{M}_{\odot})$ &  0.1 & 0.32 & 1.0 \\
    $\macc$ & $(\mathrm{M}_{\odot}\ \mathrm{yr}^{-1})$ &  $4\times10^{-11}$ & $2\times10^{-9}$ &  $1\times10^{-8}$ \\  
    \hline 
    & & \multicolumn{3}{c}{$\alpha_{\rm visc} = 10^{-2}$}\\
    \hline
    \tvisc & ($\times10^{6}\ \mathrm{yr}$) & 0.035 & 0.046 & 0.115 \\
    \minit & $(\mathrm{M}_{\odot})$ & $4.5\times10^{-4}$ & $1.99\times10^{-2}$ & $6.9\times10^{-2}$ \\
    \hline
    & & \multicolumn{3}{c}{$\alpha_{\rm visc} = 10^{-3}$}\\
    \hline
    \tvisc & ($\times10^{6}\ \mathrm{yr}$) & 0.35 & 0.46 & 1.15 \\
    \minit & $(\mathrm{M}_{\odot})$ & $2.1\times10^{-4}$ & $1.0\times10^{-2}$ & $5.9\times10^{-2}$ \\
    \hline
    & & \multicolumn{3}{c}{$\alpha_{\rm visc} = 10^{-4}$}\\
    \hline
    \tvisc & ($\times10^{6}\ \mathrm{yr}$) & 3.5 & 4.6 & 11.5 \\
    \minit & $(\mathrm{M}_{\odot})$ & $4.1\times10^{-4}$ & $2.5\times10^{-2}$ & $2.6\times10^{-1}$ \\
    \hline
    \end{tabular}
\end{table}

\subsection{The {\normalfont \texttt{DALI}} models}
\label{sec: DALI models}

Based on our 9 sets of initial conditions, we calculate \mdisk\ and \rc\ at 10 disk ages between 0.1 and 10 Myr using equations \eqref{eq: mass time evolution} and (3) (see Figure \ref{fig: disk parameter evolution} for an example). From \mdisk(t) and \rc(t), we calculated $\Sigma_{\rm gas}$(t) and use that as input for the thermochemical code \underline{D}ust \underline{a}nd \underline{Li}nes (\texttt{DALI}; \citealt{Bruderer2012,Bruderer2013}).
Based on a 2D physical disk structure, \texttt{DALI} calculates the thermal and chemical structure of the disk self-consistently. First, the dust temperature structure and the internal radiation field are computed using a 2D Monte Carlo method to solve the radiative transfer equation.
In order to find a self-consistent solution, the code iteratively solves the time-dependent chemistry, calculates molecular and atomic excitation levels and computes the gas temperature by balancing heating and cooling processes. The model can then be ray-traced to construct synthetic emission maps. 
For a more detailed description of the code we refer the reader to Appendix A of \cite{Bruderer2012}.

For the vertical structure of our models we assume a Gaussian density distribution, with a radially increasing scale height of the form $h = h_c \left(R/R_{\rm c}\right)^{\psi}$. Here $h_c$ is the scale height at $R_{\rm c}$ and $\psi$ is the flaring angle.
The stellar spectrum used in our models is a blackbody with $T_{\rm eff} = 4000$ K. To this blackbody we add excess UV radiation, resulting from accretion, in the form of 10000 K blackbody. For the luminosity of this component, we assume that the gravitational potential energy of the accreted mass is released with 100\% efficiency (see, e.g. \citealt{kama2015}). For the external UV radiation we assume a standard interstellar radiation field of 1 G$_0$   \citep{Habing1968}). These parameters are summarized in Table \ref{tab: model fixed parameters}.

In our models we include the effects of dust settling by subdividing our grains into two populations. A population of small grains (0.005-1 $\mu$m) follows the gas density distribution both radially and vertically. A second population of large grains (1-$10^3\ \mu$m), making up 90\% of the dust by mass, follows the gas radially but has its scale height reduced by a factor $\chi = 0.2$ with respect to the gas. We compute the dust opacities for both populations using a standard ISM dust composition following \cite{WeingartnerDraine2001}, with a MRN \citep{mathis1977} grain size distribution. 
We do not include any radial drift or radially varying grain growth in our \texttt{DALI} models (cf. \citealt{Facchini2017}). However, note that \cite{Trapman2019a} showed that dust evolution does not affected measured gas outer radii.

\begin{table}[htb]
  \centering   
  \caption{\label{tab: model fixed parameters}Fixed \texttt{DALI} parameters of the physical model.}
  \begin{tabular*}{0.8\columnwidth}{ll}
    \hline\hline
    Parameter & Range\\
    \hline
     \textit{Chemistry}&\\
     Chemical age & 0.1-10$^{*,\dagger}$ Myr\\
     {[C]/[H]} & $1.35\cdot10^{-4}$\\
     {[O]/[H]} & $2.88\cdot10^{-4}$\\
     \textit{Physical structure} &\\ 
     $\gamma$ &  1.0\\ 
     $\psi$ & 0.15\\ 
     $h_c$ &  0.1 \\ 
     \rc & $10-3\times10^{3,\dagger}$ AU\\
     $M_{\mathrm{gas}}$ & $10^{-5} - 10^{-1,\dagger}$ M$_{\odot}$ \\
     Gas-to-dust ratio & 100 \\
     \textit{Dust properties} & \\
     $f_{\mathrm{large}}$ & 0.9 \\
     $\chi$ & 0.2 \\
     composition & standard ISM$^{1}$\\
     \textit{Stellar spectrum} & \\
     $T_{\rm eff}$ & 4000 K + Accretion UV \\
     $L_{*}$ & 0.28 L$_{\odot}$  \\
     $\zeta_{\rm cr}$ & $10^{-17}\ \mathrm{s}^{-1}$\\
     \textit{Observational geometry}&\\
     $i$ & 0$^{\circ}$ \\
     PA & 0$^{\circ}$ \\
     $d$ & 150 pc\\
    \hline
  \end{tabular*}
  \captionsetup{width=.75\columnwidth}
  \caption*{\footnotesize{$^*$The age of the disk is taken into account when running the time-dependent chemistry. $^{\dagger}$These parameters evolve with time (see Figure \ref{fig: disk parameter evolution} and Sect. \ref{sec: effect of CO depletion}).
  $^{1}$\citealt{WeingartnerDraine2001}, see also Section 2.5 in \citealt{Facchini2017}. }}
\end{table}

\section{Results}
\label{sec: results}

\subsection{Time evolution of the $^{12}$CO emission profile}
\label{sec: viscous evolution of CO profile}

To first order, the evolution of $^{12}$CO intensity profile is determined by three time-dependent processes:
\begin{enumerate}
        \item Viscous spreading moves material, including CO, to larger radii resulting in more extended CO emission.
        \item The disk mass decreases with time, lowering surface density, which in the outer disk allows CO to be more easily photo-dissociated. This removes CO from the outer disk and lowers the CO flux coming from these regions.
        \item Over longer timescales, time dependent chemistry will result in CO being converted into CH$_4$, CO$_2$ and CH$_3$OH. 
        This is discussed separately in more detail in Section \ref{sec: effect of CO depletion}.  
\end{enumerate}

The combined effect of the first two processes on the $^{12}$CO emission profile can be seen in Figure \ref{fig: 12CO profile evolution} for disks with a stellar mass of $M_* = 0.1\ \mathrm{M}_{\odot}$. 
Similar profiles for the remaining disks are shown in Figure \ref{fig: 12CO profiles non-depleted models}

For a high \alp\ of 0.01 the viscous timescale is short compared to the disk age and viscous evolution is happening fast. This is reflected in the $^{12}$CO emission, which spreads quickly (within 1 Myr) from $\sim200$ AU to $\sim400$ AU. After $\sim2$ Myr the $^{12}$CO emission in the outer parts of the disk starts to decrease. At this point the total column densities in the outer disk are low enough that CO is removed through photodissociation. 
As a reference, by 2 Myr the disk mass of the models has dropped to $\mdisk=5\times10^{-5}\ \msun$ and its characteristic size has increased to $\rc = 400$ AU (see Figure \ref{fig: disk parameter evolution}). 

For models with $\alp = 10^{-3}$, shown in the middle panel of Figure \ref{fig: 12CO profile evolution}, viscous spreading of the disk dominates the evolution of the $^{12}$CO emission profile. Compared to the $\alp = 10^{-2}$ models the column density in the outer disk never becomes low enough for CO to be efficiently photo-dissociated. 

The models with $\alp = 10^{-4}$, presented in the right panel of Figure \ref{fig: 12CO profile evolution}, shows only small changes in the emission profile. For these models the viscous timescale is $\sim 3.5$ Myr, meaning that within the 10 Myr lifetime considered the surface density does not go through much viscous evolution. 

\subsection{Evolution of the observed gas outer radius}
\label{sec: gas evolution}

\begin{figure}[htb]
    \centering
    \includegraphics[width=\columnwidth,clip,trim={0cm 0.1cm 0cm 0.1cm}]{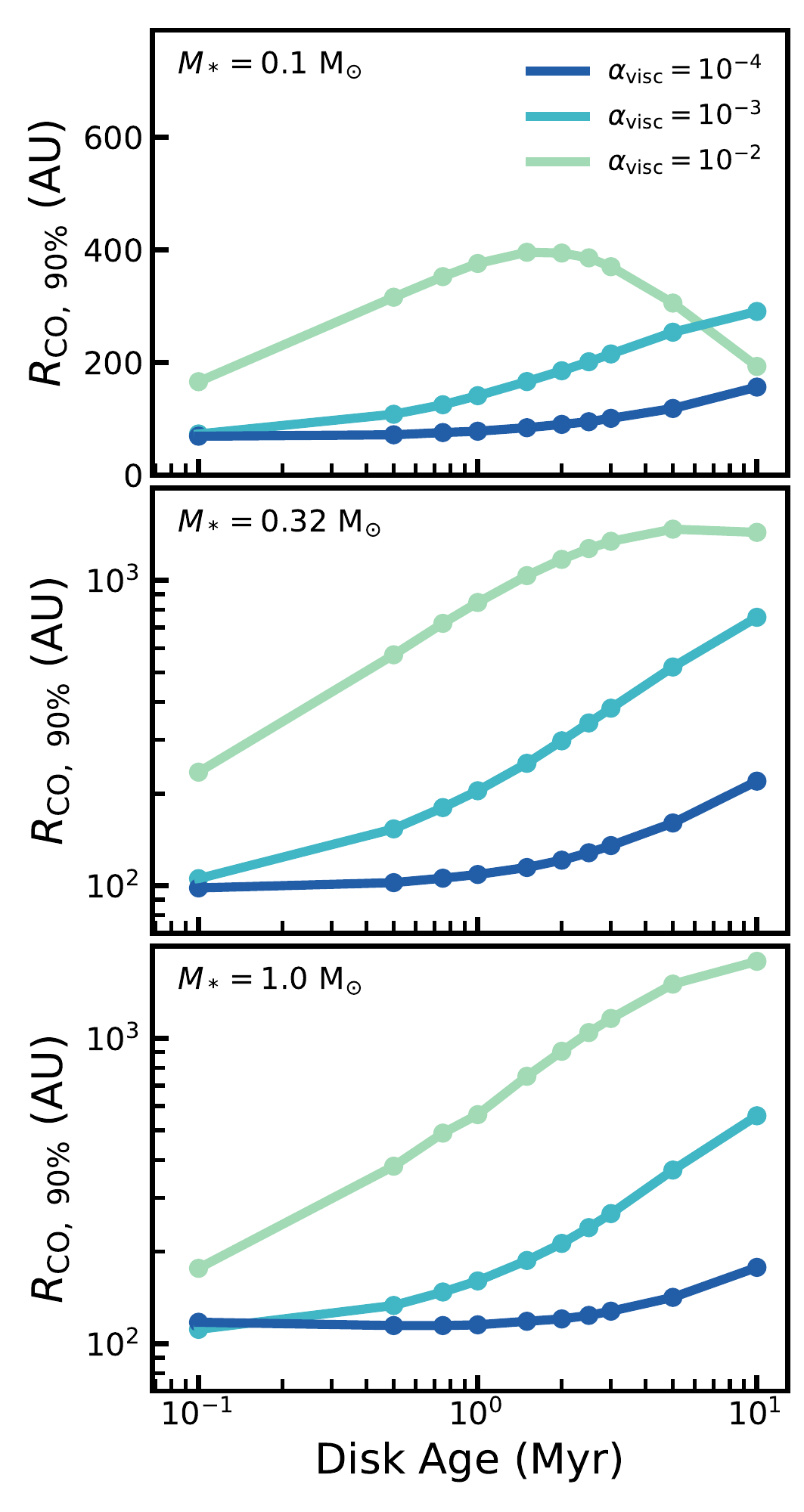}
    \caption{\label{fig: rgas evolution} Gas outer radius vs. disk age, here defined as the radius that encloses 90\% of the $^{12}$CO 2\,-\,1 flux. Top, middle and bottom panels show models with different stellar masses (cf. Table \ref{tab: model initial conditions}). Colors correspond to the \alp\ of the model.}
\end{figure}

From the $^{12}$CO emission maps we can calculate the outer radius that would be obtained from observations. We define the observed gas outer radius, \rgas, as the radius that encloses 90\% of the total $^{12}$CO flux.
A gas outer radius defined this way encloses most $(>98\%)$ of the disk mass and traces a fixed surface density in the outer disk \citep{Trapman2019a}. Note that we do not including observational factors, like noise, which affect the \rgas\ that is measured. 
To accurately retrieve \rgas\ from observations requires a peak signal-to-noise of $\mathrm{S/N} > 10$ on the moment zero map of the $^{12}$CO emission (cf. \citealt{Trapman2019a}).
Note that the radii discussed here are measured from $^{12}$CO $J=2\,-\,1$ emission, but tests show that gas outer radii measured from $^{12}$CO $J=3\,-\,2$ are the same to within a few percent.

Figure \ref{fig: rgas evolution} shows how the observed outer radius changes as a result of viscous evolution. The top panel shows \rgas\ for models with $\mstar = 0.1\ \msun$. For $\alp = 10^{-2}$ the gas outer radius first increases until at $\sim 2$ Myr it starts to decrease. The decrease in \rgas\ is due to decreasing column densities in the outer disk, allowing CO to be more easily photo-dissociated (for details, see Section \ref{sec: viscous evolution of CO profile}). 
For $\alp = 10^{-3}$, \rgas\ increases monotonically from $\sim70$ AU to $\sim280$ AU. The trend is similar for $\alp  = 10^{-4}$ but \rgas\ increases at a slower rate, ending up at $\rgas \sim 150$ AU after 10 Myr.

For models with $\mstar = 0.32\ \msun$ and $1.0\ \msun$, the initial and final disk masses are much higher compared to the models with $\mstar = 0.1\ \msun$. As a result, photo-dissociation does not have a significant effect on \rgas\ and \rgas\ does not significantly decrease with age.
In addition, the disk sizes for these two groups of models are very similar. For $\alp = 10^{-2}$, \rgas\ instead rapidly increases from $\sim180-250$ AU at 0.1 Myr to $1500-1800$ AU at 10 Myr. For $\alp = 10^{-3}$ the growth of \rgas\ is less extreme in comparison, but observed disk sizes still reach $500-700$ AU after 10 Myr. Due to the long viscous timescales of $5-10$ Myr for the models with $\alp = 10^{-4}$, \rgas\ does not increase significantly, i.e. by less than a factor of $\sim2$, over a disk lifetime of $\sim10$ Myr. 

For more embedded star forming regions the $^{12}$CO emission from the disk can be contaminated by the cloud, either by having $^{12}$CO emission from the cloud mixed in with the emission from the disk or through cloud material along the line of sight absorbing the $^{12}$CO emission from the disk.
This can prevent using $^{12}$CO to accurately measure disk sizes. 
We have therefore also examined disk sizes measured using 90\% of the $^{13}$CO 2\,-\,1 flux, which are shown in Figure \ref{fig: 13CO outer radius evolution}. Apart from being a factor $\sim1.4-2$ smaller than the $^{12}$CO outer radii, the $^{13}$CO outer radii evolve similarly as \rgas. The $^{13}$CO outer radii are smaller than \rgas\ because the less abundant $^{13}$CO is more easily removed in the outer parts of the disk through photo-dissociation.
Our conclusions for \rgas\ are therefore also applicable for gas disk sizes measured from $^{13}$CO emission. However, see Section \ref{sec: effect of CO depletion} on how chemical depletion of CO through grain-surface chemistry affects this picture.

Overall we find that the observed outer radius increases with time and is larger for a disk with a higher \alp. The exception to this rule are the models with low stellar mass $(M_* = 0.1\ \mathrm{M}_{\odot})$ and high viscosity $(\alp = 10^{-2})$. They highlight the caveat
that if the disk mass becomes too low, CO becomes photo-dissociated in the outer disk and the observed outer radius will decrease with time. 

\subsection{Is the gas outer radius tracing viscous spreading?}
\label{sec: rgas vs rc}

\begin{figure}[htb]
    \centering
    \includegraphics[width=0.97\columnwidth,clip,trim={0cm 0.3cm 0cm 0.1cm}]{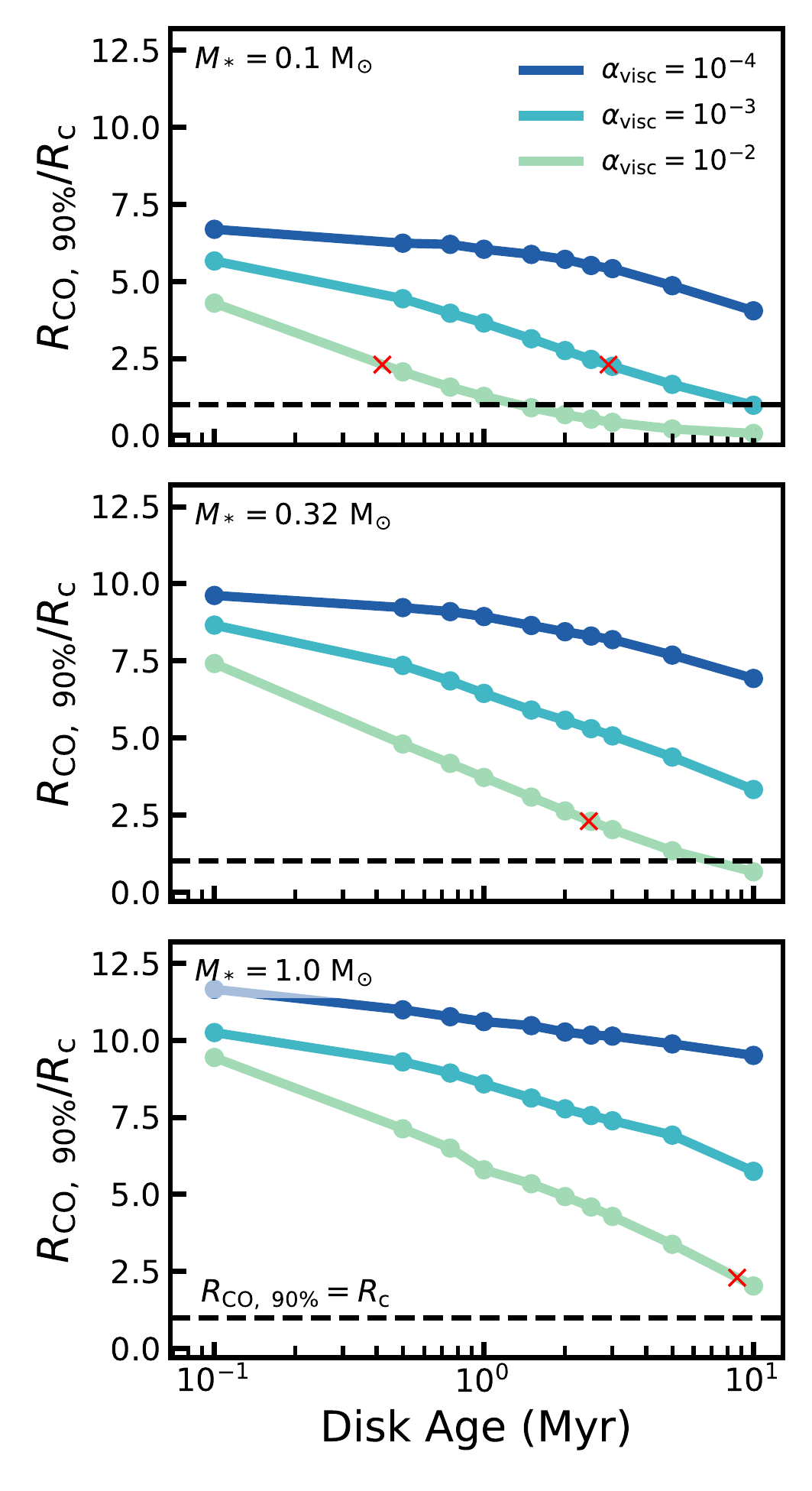}
    \caption{\label{fig: rgas traces rc} The ratio of gas outer radius over characteristic radius vs. disk age. The black dashed line shows where $\rgas = \rc$. 
    Top, middle and bottom panels show models with different stellar masses (cf. Table \ref{tab: model initial conditions}). Colors correspond to the \alp\ of the model. Red crosses denote where $\rgas/\rc = 2.3$, i.e., where \rgas\ encloses 90\% of the total disk mass (see Section \ref{sec: retrieving alpha})}
\end{figure}

The previous section has shown that disks with higher \alp, which evolve over a shorter viscous timescale, are overall larger at a given disk age. To quantify this, it is worthwhile to examine how well the observed gas outer radius \rgas\ traces the characteristic size \rc\ of the disk.

Figure \ref{fig: rgas traces rc} shows the ratio $\rgas/\rc$ as function of disk age for the three sets of stellar masses. If \rgas\ were only affected by viscous spreading it would grow at the same rate as the characteristic radius, $\rgas \propto \rc$, represented by a horizontal line in Figure \ref{fig: rgas traces rc}. Instead we see $\rgas/\rc$ decreasing with disk age, indicating that the observed outer radius grows at a slower rate than the viscous spreading of the disk. 
The main cause for the slower growth rate of \rgas\ is the decreasing disk mass over time, due to the fact that \rgas\ traces a fixed surface density. As shown in \cite{Trapman2019a}, \rgas\ coincides with the location in the outer disk where CO is no longer able to effectively self-shield against photodissociation and is quickly removed from the gas. The CO column density threshold for CO to self-shield is $N_{\rm CO} \geq 10^{15}\ \mathrm{cm}^{-2}$ \citep{vanDishoeckBlack1988}. Thus, given that \rgas\ traces a point of fixed column density, it scales with the total disk mass. As a result of angular momentum transport via viscous stresses, material is accreted onto the star, causing the total disk mass to decrease following equation \eqref{eq: mass time evolution}  (see, e.g, Figure \ref{fig: disk parameter evolution}), which limits the growth of \rgas.

Figure \ref{fig: rgas traces rc} also shows that $\rgas/\rc$ is larger for models with a lower \alp, with this difference becoming larger for older disks. This behavior can also be related to disk mass. As shown in Figure \ref{fig: disk parameter evolution} disk models with a lower \alp\ have a higher disk mass. 
For the same \rc\ a higher disk mass will, to first order, result in higher CO column densities in the outer disk. As a result CO is able to self-shield against photo-dissociation further out in the disk, increasing the difference between \rgas\ and \rc. 

In conclusion, we find that $\rgas/\rc$ is between 0.1 and 12 and is mainly determined by the time evolution of the disk mass, which is set by the assumed viscosity. To infer \rc\ directly from \rgas\ requires information on the total disk gas mass.

\begin{figure}[thb]
    \centering
    \includegraphics[width=0.93\columnwidth,clip,trim={0cm 0.2cm 0.3cm 0.7cm}]{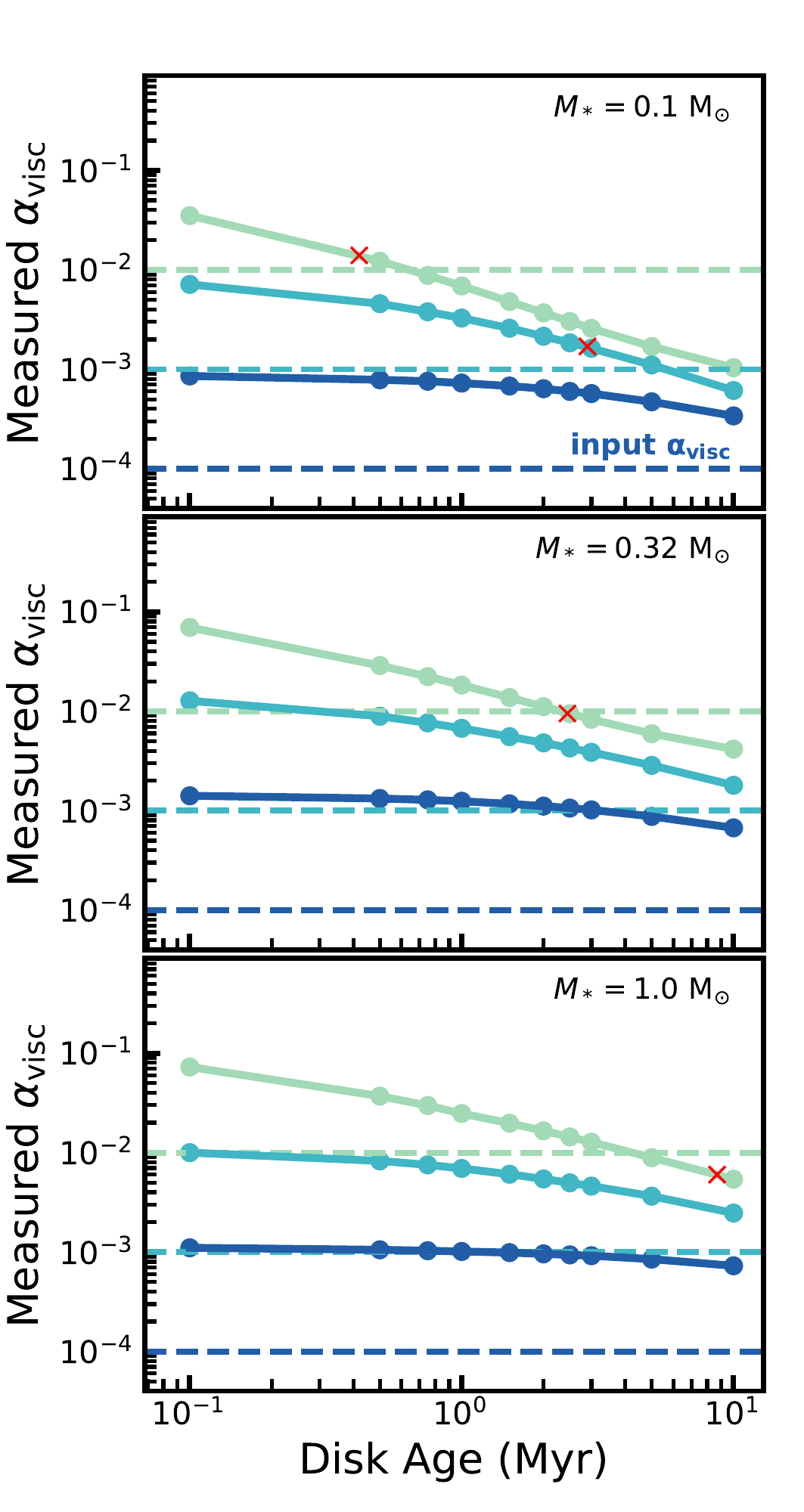}
    \caption{\label{fig: retrieving alpha} Comparison between \alp\ measured from \rgas\ (solid line) and the input \alp\ (dashed line). Colors correspond to the input \alp\ of the model. Top, middle and bottom panels show models with different stellar masses (cf. Table \ref{tab: model initial conditions}). Red crosses denote where $\rgas/\rc = 2.3$ and \rgas\ encloses 90\% of \mdisk\ (cf. Section \ref{sec: retrieving alpha} and Figure \ref{fig: rgas traces rc}). Red crosses denote where $\rgas/\rc = 2.3$, i.e., where \rgas\ encloses 90\% of the total disk mass (see Section \ref{sec: retrieving alpha}).} 
\end{figure}

\subsection{How well can \alp\ be measured from observed \rgas?}
\label{sec: retrieving alpha}

A useful definition for an outer radius of a disk is that it encloses most of the mass in the disk. In this case, using a few simple assumptions, we can relate the outer radius directly to \alp. If we assume that the viscous timescale at the outer radius of the disk, given by $\tvisc \approx R_{\rm out}^2\nu^{-1}$, is approximately equal to the age of the disk, given by $\macc \approx \mdisk/\tvisc$, we can write \alp\ as (see, e.g., \citealt{Hartmann1998,Jones2012,Rosotti2017})
\begin{equation}
\label{eq: alpha derivation}
    \alp = \frac{\macc}{\mdisk} \cdot c_s^{-2} \cdot \Omega_K \cdot R_{\rm out}^2
\end{equation}
where \macc\ is the stellar accretion rate, \mdisk\ is the total disk mass, $c_s$ is the sound speed, $\Omega_K$ is the Keplerian orbital frequency and $R_{\rm out}$ is the physical outer radius of the disk. 

In absence of this physical outer radius, \cite{ansdell2018} used the observed gas outer radius \rgas, based on the $^{12}$CO 2-1 emission, to measure \alp\ for 17 disks in Lupus finding a wide range of \alp, spanning two orders of magnitude between $3\times10^{-4}$ and $0.09$.

For our models we have both \rgas, measured from our models, and the input \alp, so we can examine how well \alp\ can be retrieved from the observed gas outer radius \rgas. As we are mainly interested in the correlation between \alp\ and \rgas, we will assume that \mstar, \macc\ and \mdisk\ are known and $c_s$ is calculated assuming a disk temperature of 20 K, the same temperature used to calculate $t_{\rm visc}$ in Section \ref{sec: initial condition}. 

Figure \ref{fig: retrieving alpha} shows \alp\ measured using \rgas, for our models and compares it to the \alp\ that was used as input for the models. For all disk models the measured \alp\ decreases with age and, for most ages, we find \alp(measured)$\,>\,$\alp(input). Both of these observations can be traced back to which radius is used in Equation \eqref{eq: alpha derivation} to calculate \alp(measured).
In the assumptions going into deriving Equation \eqref{eq: alpha derivation}, $R_{\rm out}$ is defined as the radius that encloses all (100 \%) of the mass of the disk. In our models where the surface density follows a tapered power law the radius that encloses 100 \% of the mass is infinite, but we can instead take $R_{\rm out}$ as a radius that encloses a large, fixed fraction of the disk mass. For a tapered powerlaw this $R_{\rm out}$ is directly related to \rc. As an example, for a tapered powerlaw with $\gamma = 1$, the radius that encloses 90\% of the disk mass is $2.3\times\rc$. For $R_{\rm out} = 2.3\times\rc$ in Equation \eqref{eq: alpha derivation} we obtain approximately the same \alp\ that was put into the model. Ideally we would therefore like \rgas\ to also enclose a large, fixed fraction of the disk mass, or continuing our example, we would like $\rgas \approx 2.3\times\rc$, independent of disk age and mass. 
However, in Section \ref{sec: rgas vs rc} we have shown that \rgas/\rc\ lies between 0.1 and 10 and decreases with disk age. Figure \ref{fig: rgas traces rc} shows that $\rgas/\rc \gg 2.3$ for most disk ages, leading us to overestimate \alp\ when measured from \rgas. Taking the example discussed before, if we compare Figures \ref{fig: retrieving alpha} and \ref{fig: rgas traces rc} we find that at disk ages where $\rgas/\rc\approx2.3$ our \alp\ measured from \rgas\ is within a factor of 2 of the input \alp.

Summarizing we find that in most cases, $\rgas/\rc \gg 2.3$ and we measure an \alp\ much larger than the input \alp, up to an order of magnitude higher, especially if the input \alp\ is low. 
Given that at 1 Myr the measured \alp\ is $5-10\times$ larger than the input \alp, this implies that the \alp\ determined by \cite{ansdell2018} likely overestimates the true \alp\ of the disks in Lupus by a factor 5-10. 

\section{Discussion}
\label{sec: discussion}
\begin{figure}[htb]
    \centering
    \includegraphics[width=\columnwidth]{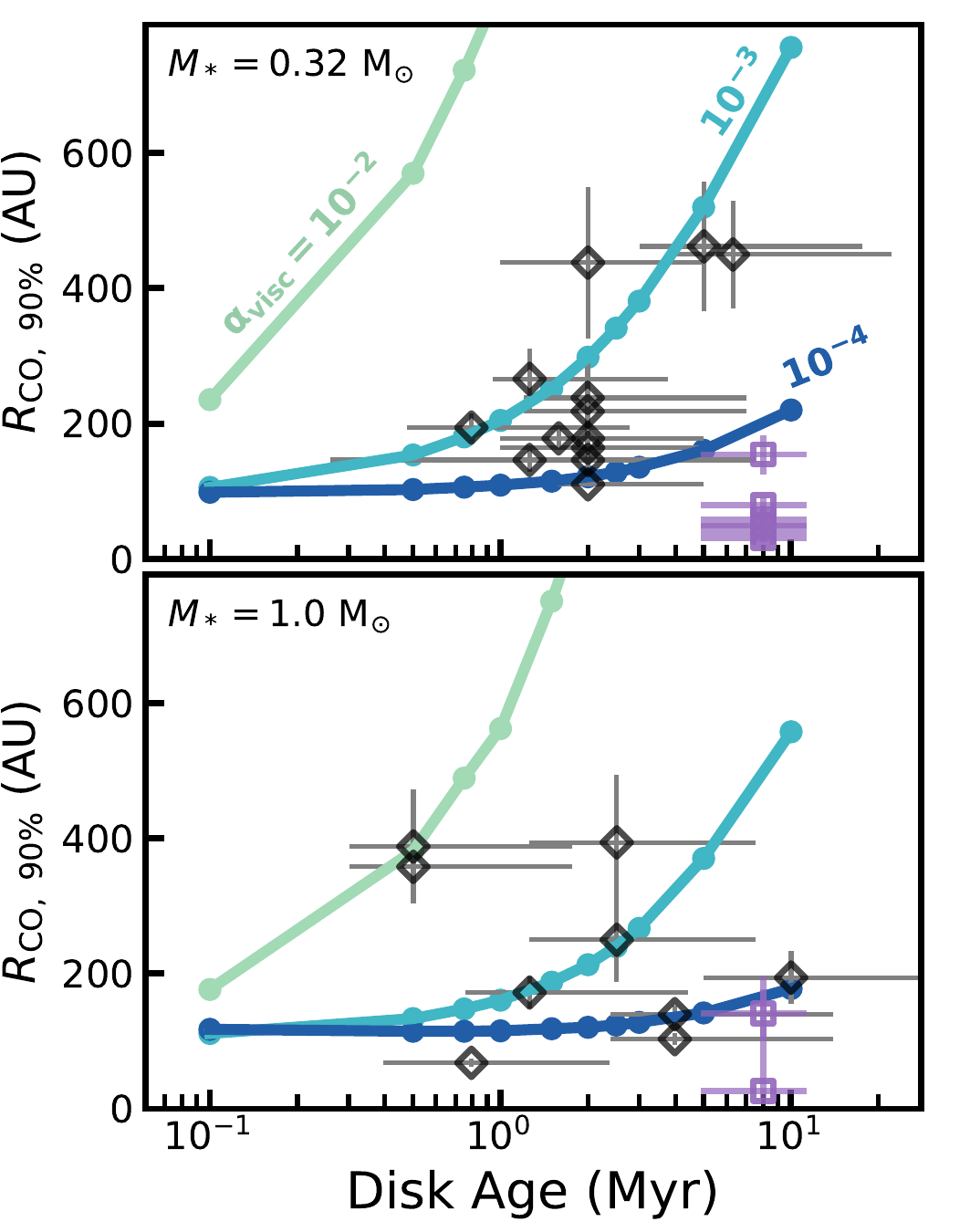}
    \caption{\label{fig: rgas evolution w observations} 
    Gas outer radii of our models (\rgas) compared to observations. Colours correspond to the \alp\ of the model.
    Black open diamonds show observed gas outer radii in Lupus \citep{ansdell2018} and purple open squares denote observed gas outer radii in Upper Sco \citep{barenfeld2017}. 
    Note that the  Upper Sco outer radii shown here are 90\% outer radii, calculated from their fit to the observed $^{12}$CO intensity.
    Top and bottom panels split models and observations based on stellar mass. Sources with $\mstar \leq 0.66\ \msun$ are compared to models with $\mstar = 0.32\ \msun$, those with $\mstar > 0.66\ \msun$ are compared to models with $\mstar = 1.0\ \msun$. 
    We only show panels for $\mstar = 0.32$ and $1.0\ \mathrm{M}_{\odot}$ as the sample of observations considered here does not contain any objects with $\mstar\sim 0.1\ \mathrm{M}_{\odot}$. 
    } 
\end{figure}

\subsection{Comparing to observations}
\label{sec: comparing to observations}

Gas disk sizes have now been measured consistently for a significant number of disks.
In contrast with \cite{NajitaBergin2018}, in this paper we have chosen to select homogeneous samples (in terms of analysis and tracer). These samples are \cite{ansdell2018} and \cite{barenfeld2017}, who measured \rgas\ for 22 sources in Lupus and for 7 sources in Upper Sco. Between them, these disks span between 0.5 and 11 Myr in disk ages. In Figure \ref{fig: rgas evolution w observations} we compare our models to these observations, where sources with $\mstar \leq 0.66\ \msun$ are compared to models with $\mstar = 0.32\ \msun$ and those with $\mstar > 0.66\ \msun$ are compared to models with $\mstar = 1.0\ \msun$. \cite{ansdell2018} has defined the gas outer radius as the radius that encloses 90\% of the $^{12}$CO 2\,-\,1 emission, so we take directly their values.
For the disks in Upper Sco we calculate \rgas\ from their fit to the observed $^{12}$CO intensity (cf. \citealt{barenfeld2017}).
Stellar ages and masses were determined by comparing pre-main sequence evolutionary models to X-shooter observations of these sources \citep{alcala2014,alcala2017}. 
Lacking such observations for Upper Sco, we instead use the 5-11 Myr stellar age of Upper Sco (see, e.g. \citealt{Preibisch2002,Pecaut2012}) for all sources in this region. The observations are summarized in Table \ref{tab: observations}.

As shown in Figure \ref{fig: rgas evolution w observations}, most of the Lupus observations lie between the models with $\alp = 10^{-3}$ and $\alp = 10^{-4}$. Most of the disks can therefore been explained as viscously spreading disks with $\alp = 10^{-4} - 10^{-3}$ 
that start out small $(\rinit = 10\ \mathrm{AU})$. 
Only two sources with $\mstar = 1.0\ \msun$, IM Lup  and Sz 98 (also known as HK Lup), require a larger $\alp \simeq 10^{-2}$ to explain the observed gas disk size given their age.

It is interesting to note that \cite{Lodato2017} reached a similar conclusion 
using a completely different method. They show that a simple viscous model can reproduce the observed relation between stellar mass accretion and disk mass in Lupus (see \citealt{manara2016b}). To match both the average disk lifetime and the observed scatter in the \macc-\mdisk\ relation,  disk ages in Lupus have to be comparable to the viscous timescale, on the order of $\sim 1$ Myr (see also \citealt{Jones2012,Rosotti2017}). This viscous timescale is comparable to our models with $\alp = 10^{-3}$ (cf. Table \ref{tab: model initial conditions}).

As we have made no attempt to match our models to individual observations, it is worthwhile to discuss if it is possible to explain the large disks in our sample, like IM Lup and Sz 98, by other means than a large \alp. A quick comparison with Table \ref{tab: model initial conditions} shows that increasing the disk mass will not help to explain their large \rgas. Our models with $\mstar=1.0\ \msun$ and $\alp=10^{-3} - 10^{-4}$ differ by an order of magnitude in initial disk mass, but at 0.75 Myr they differ by less than 5\% in terms of \rgas. Another possibility would be increasing the initial disk size $\rc(t=0)$, which we discuss in Section \ref{sec: larger initial disk sizes}.

Interestingly, 5 of the 7 Upper Sco disks have gas outer radii that lie well below our models with $\alp = 10^{-4}$, indicating that their small \rgas\ cannot be explained by our models, even when taking into account the uncertainty on their age. As the viscous timescale for our models with $\alp = 10^{-4}$ is already $\sim10$ Myr, decreasing \alp\ will not allow us to reproduce the observed \rgas. At face value, these small disk sizes would thus seem to rule out that these disks have evolved viscously.
Note, however, that there are processes which, in combination with viscous evolution, could explain these small disk sizes.
The first would consist in reducing the CO content of these disks; we discuss this option in Section \ref{sec: effect of CO depletion}. Given that the disks in Upper Sco are highly irradiated as members of the Sco-Cen OB association (see e.g. Sections \ref{sec: effects of photo-evaporation} and Appendix \ref{app: upper sco irradiation}), another option is that external photo-evaporation is the culprit for their small disk sizes, but we cannot rule out a contribution of MHD disks winds to their evolution.

We have shown that the current observations in Lupus are consistent with viscous disk evolution with a low effective viscosity of $\alp = 10^{-3} - 10^{-4}$ (viscous timescales of $1-10$ Myr). However, the current available data do not provide sufficient evidence for viscous spreading, the proof that viscosity is driving disk evolution. We stress that this is mostly because most of the available data come from the same star-forming region (Lupus), and therefore most of the disks are concentrated around a disk age range of 1-3 Myr. Only a few disks lie outside this range. 
The inclusion of the Upper Sco disks would help constrain the importance of viscous spreading,
but we have already commented in the previous paragraph about the caveats with this region. Thus, we conclude that the current sample of radii is insufficient to confirm or reject the hypothesis that disks are viscously spreading. To overcome this problem, future observational campaigns should focus on expanding the observational sample of well measured disk CO radii in other star forming regions with a different age than Lupus. 

\subsection{Larger initial disk sizes}
\label{sec: larger initial disk sizes}

\begin{figure}
    \centering
    \includegraphics[width=\columnwidth]{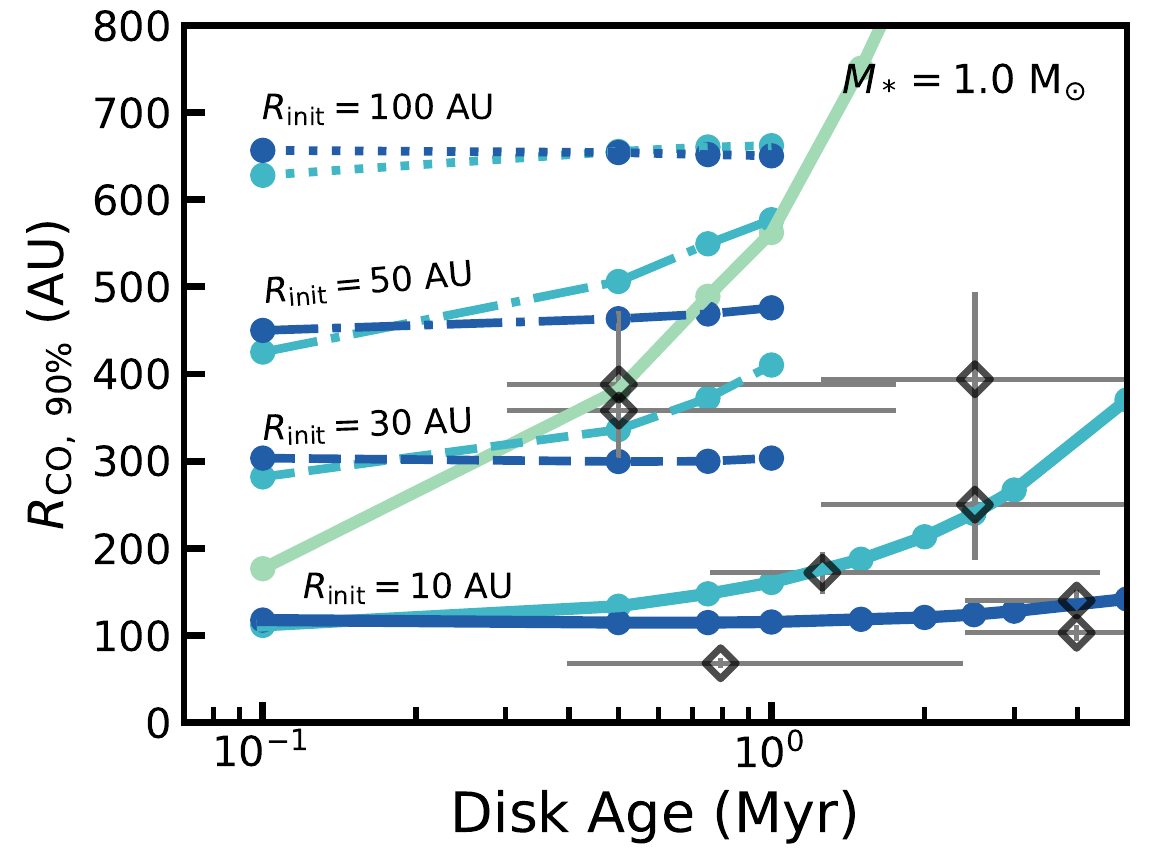}
    \caption{\label{fig: larger initial disk sizes} Zoom in of the bottom panel of Figure \ref{fig: rgas evolution w observations}, showing gas outer radii of our models (\rgas) compared to observations. Added in here are models with $\mstar = 1.0~\msun$ (see Section \ref{sec: initial condition}) but with $\rinit = 30, 50$ and 100 AU, denoted with dotted, dashed-dotted and dashed lines, respectively. Colors indicated different values for \alp. Note that the new models were only run for $\alp = [10^{-3}, 10^{-4}]$.}
\end{figure}

In our analysis we have assumed that disks start out small, with $\rinit = 10$ AU, as motivated by recent ALMA observations of disks around young Class 0 and 1 protostars \citep{Maury2019,Maret2020,Tobin2020}. However, these observations also show a spread in disk size for these young objects. Increasing the initial disk size would potentially also let us explain the larger disks, for example IM Lup and Sz 98 (see \citealt{vanTerwisga2018}), with a lower \alp.

Figure \ref{fig: larger initial disk sizes} presents \rgas\ measured from three sets of models with $\mstar=1.0\ \msun$ (see Section \ref{sec: initial condition}), but with an increased $\rinit = [30, 50, 100]$ AU. Since our models with $\rinit = 10$ AU and $\alp = 10^{-2}$ already have \rgas\ much larger than what is observed, we run these new models only for $\alp = [10^{-3}, 10^{-4}]$. These models have much larger gas disk sizes than models with $\rinit=10$ AU, with \rgas\ being at least 3 times larger ($\rgas \geq 300$ AU). As such, they show that the large disks in the sample ($R_{\rm CO,\ 90\%, obs.}\geq 300\ \mathrm{AU}$) can also be explained with a larger initial size $(\rinit = 30)$ and a low viscous alpha $(\alp = 10^{-4})$. 
Extrapolating our results here beyond 1 Myr and to models with $\mstar=0.32\ \msun$ there are six of such large disks in Lupus that can be explained with $\rinit \approx 30$ AU (c.f. Figure \ref{fig: rgas evolution w observations}).
In particular these models show that the observed gas disk sizes of IM Lup and Sz 98 can be explained by either a high viscous alpha $(\rinit = 10\ \mathrm{AU}, \alp = 10^{-2})$ or a larger initial disk size $(\rinit \simeq 30-50\ \mathrm{AU}, \alp = 10^{-3}-10^{-4})$.
Given the similarities in terms of \rgas\ between models with $\mstar=1.0\ \msun$ and $\mstar=0.32\ \msun$ seen in Figure \ref{fig: rgas evolution}, we expect that increasing \rinit\ from 10 AU to 30 AU for models with $\mstar=0.32\ \msun$ would similarly increase their \rgas\ by a factor of at least 3.

However, our models show that disks with $\rinit = 30$ AU start out at $\rgas\sim300$ AU, which is already much larger than most observed \rgas\ in Lupus \citep{ansdell2018}. This indicates that, even if a larger \rinit\ can provide an explanation for these six disks, the bulk of the disks in Lupus cannot have had a large \rinit\ and  they must have started out small $(\rinit \simeq 10\ \mathrm{AU})$.
This could be in line with the observations of \cite{Tobin2020} although their measured dust radii should be multiplied by a factor of typically $2-3$ to get the gas radii due to optical depth effects \citep{Trapman2019a}.

\begin{figure*}[thb]
    \centering
    \includegraphics[width=0.9\textwidth]{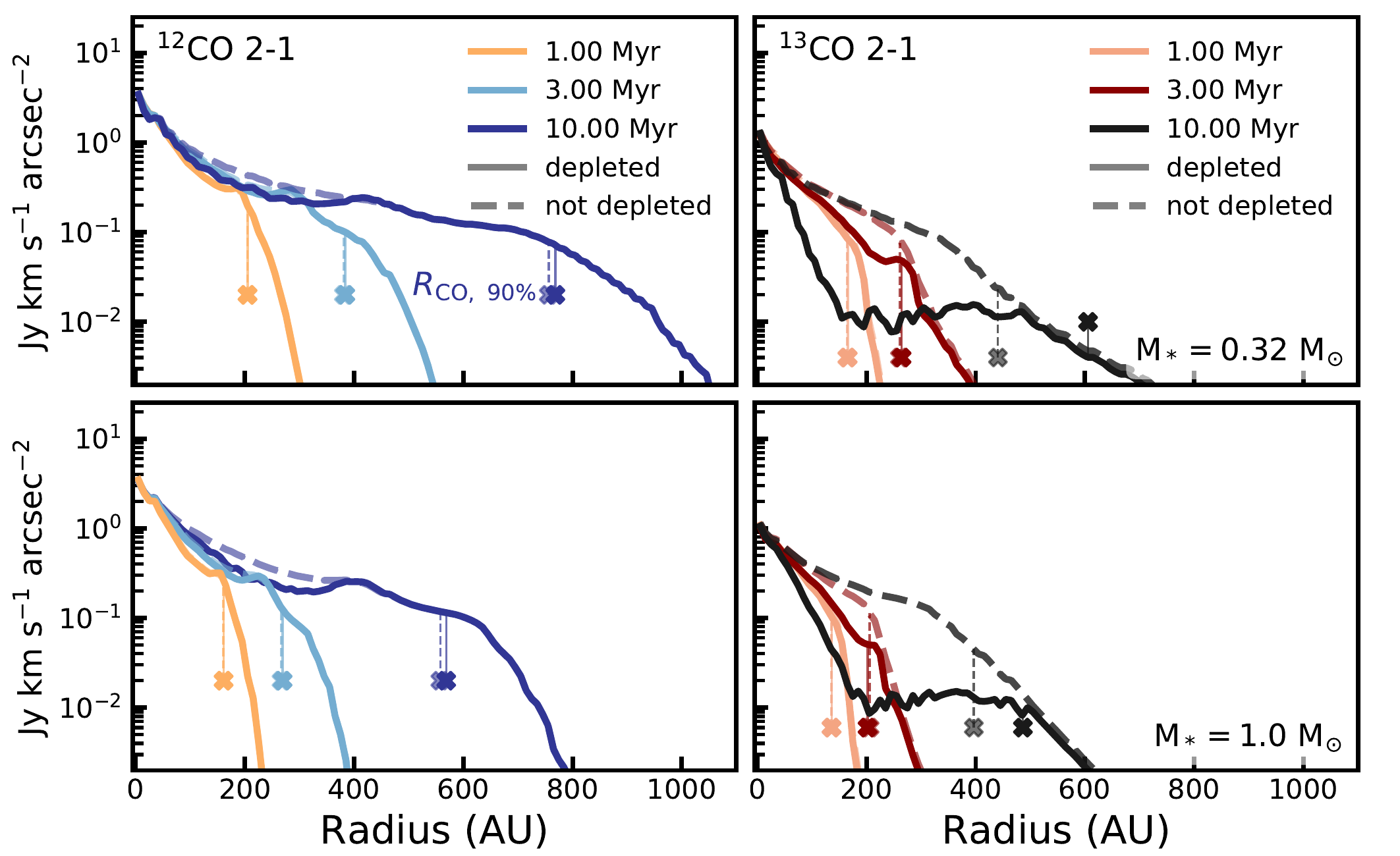}
    \caption{\label{fig: CO depleted intensity profiles} The effect of chemical CO depletion through grain-surface chemistry on the $^{12}$CO 2\,-\,1 intensity profile (left panels) and $^{13}$CO 2\,-\,1 intensity profile (right panels) after 1 Myr (orange), 3 Myr (light blue/dark red) and 10 Myr (blue/black). Top and bottom rows show models with $\mstar = 0.32\ \mathrm{M_{\odot}}$ and $\mstar = 1.0\ \mathrm{M_{\odot}}$, respectively.
    The profile without chemical CO depletion is shown as a dashed line.
    The gas outer radii (\rgas) are shown as a cross at arbitrary height below the profile.
    Note that after 1 Myr the chemical CO depletion is not significant enough to change the intensity profile and \rgas. After 10 Myr chemical CO depletion has caused \rgas\ to increase (for details, see Section \ref{sec: effect of CO depletion}).
    Figures \ref{fig: 13CO profiles non-depleted models} and \ref{fig: 13CO profiles depleted models} give a full overview of the $^{13}$CO 2\,-\,1 intensity profile of the models without and with chemical CO depletion. }
\end{figure*}

\subsection{Is chemical CO depletion affecting measurements of viscous spreading?}
\label{sec: effect of CO depletion}

Over the recent years it has become apparent in observations that protoplanetary disks are underabundant in gaseous CO with respect to the expected abundance of $\mathrm{CO/H}_2 = 10^{-4}$
(see, e.g. \citealt{Favre2013,Du2015,Kama2016,Bergin2016,Trapman2017}). Several authors have shown that grain-surface chemistry is able to lower the CO abundance in disks, by converting CO into CO$_2$ and CH$_3$OH on the grains on a timescale of several Myr (see, e.g. \citealt{Bosman2018b,Schwarz2018}). 
In this work we will refer to this process as \emph{chemical depletion of CO} to distinguish it from simple freeze out of CO which is included in our models presented in Section \ref{sec: results}.
As the chemical depletion of CO operates on similar timescales as viscous evolution, it can have a large impact on the use of  $^{12}$CO as a probe for viscous evolution. Here we implement an approximate description for grain surface chemistry and examine its effects on observed gas outer radii. A more detailed description can be found in Appendix \ref{app: implementation CO depletion}. 

Figure \ref{fig: CO depleted intensity profiles} shows $^{12}$CO 2\,-\,1 and $^{13}$CO 2\,-\,1 intensity profiles, with and without including chemical CO depletion, for two models at different disk ages. 
The $^{12}$CO 2\,-\,1 radial intensity profile remains unchanged until 10 Myr, at which point the intensities start to drop between $\sim100$ AU and $\sim300$ AU, seemingly carving a small ``dip'' in the intensity profile. With our current definition of the outer radius at 90\% of the total flux, this dip lies within \rgas. 
The decreasing intensity due to chemical CO depletion causes \rgas\ to move outward, although the change is small ($\leq 2\%$). The chemical depletion of CO does not affect the CO abundance beyond 300 AU (see, e.g. Figure \ref{fig: CO depletion in abundance}), so the $^{12}$CO 2\,-\,1 flux originating from $> 300$ AU now makes up a larger fraction of the total $^{12}$CO flux when comparing models with and without chemical CO depletion. It should therefore be noted that if we were to change our definition of the gas outer radius such that it lies within the ``dip'', e.g. by defining $R_{\rm CO,\ X\%}$ using a lower percentage of the total flux, $R_{\rm CO,\ X\%}$ would decrease if we include chemical depletion of CO.

In contrast, the $^{13}$CO 2\,-\,1 intensity profile, shown in the right panels of Figure \ref{fig: CO depleted intensity profiles}, is significantly affected by chemical CO depletion at 10 Myr. Between $\sim$100 AU and $\sim$300 AU the $^{13}$CO intensity profile has dropped by more than an order of magnitude. Again, as an consequence of the infinite signal-to-noise (S/N) of our synthetic emission maps, the decrease in the intensity profile has moved the radius enclosing 90\% of flux outward. 
For real observations, where S/N is limited, chemical depletion of CO will instead significantly decrease the observed outer radius if measured from $^{13}$CO emission. 
Figure \ref{fig: CO depleted intensity profiles} shows that gas disk sizes measured from $^{13}$CO can only be interpreted correctly if CO depletion is taken into account in the analysis.
The figure also highlights the importance of high S/N observations when measuring gas disk sizes from $^{13}$CO emission.
Given the lack of a significant sample of observed $^{13}$CO outer radii, we do not investigate further this aspect in this paper, but note that once such a sample becomes available an analysis quantifying the effect of chemical depletion on $^{13}$CO outer radii will become possible.

As shown in Figure \ref{fig: CO depletion in abundance}, there exists a vertical gradient in CO abundances. Vertical mixing, not included in our models, would move CO rich gas from the $^{12}$CO emission layer towards the midplane and exchange it with CO-poor gas. If we were to include vertical mixing, the CO abundance in the $^{12}$CO emitting layer would decrease and thus the effect of CO depletion on \rgas\ measured from $^{12}$CO would increase. The effect would be similar to what is seen for $^{13}$CO, indicating that in this case chemical CO depletion could also affect gas disk sizes measured from $^{12}$CO emission and should thus be taken into account in the analysis (see, e.g. \citealt{Krijt2018}; \krijtprep)

\subsection{Caveats}
\label{sec: effects of photo-evaporation}

\textit{Photo-evaporation:} 
In this paper we considered a disk evolving purely under the influence of viscosity. In reality, it is well known that pure viscous evolution cannot account for the observed timescale of a few Myr on which disks disperse (see \citealt{Alexander2014} for a review). Internal photo-evaporation is commonly invoked as a mass-loss mechanism to solve this problem. Because photo-evaporation preferentially removes mass from the inner disk (a few AUs), it is unlikely to change our conclusions. We note however that some photo-evaporation models \citep{Gorti2009b} have an additional peak in the mass-loss profile at a scale of 100-200 AU, which might influence our results.

Another potential concern is the effect of external photo-evaporation, i.e. mass-loss induced by the high-energy radiation emitted by nearby massive stars. In this case, the mass-loss preferentially affects the outer part of the disk \citep{Adams2004} and might therefore have an influence on the evolution of the outer disk radius, likely moving \rgas\ inward. The importance of this effect is region-dependent. A region like Lupus is exposed to relatively low levels of irradiation (see the appendix in \citealt{Cleeves2016}) and neglecting external photo-evaporation is probably safe in this case, although the effect can still be important for the largest disks \citep{Haworth2017}. For other regions, like Upper Sco, the impact of external photo-evaporation is potentially more severe since the region is part of the nearest OB association, Sco-Cen \citep{Preibisch2008}. According to the catalog of \cite{deZeeuw1999}, the earliest spectral type in the region is B0 and there are 49 B stars in the complex, suggesting that the level of irradiation can be significantly higher than in Lupus. 
A simple calculation, outlined in Appendix \ref{app: upper sco irradiation}, suggest that the disks in Upper Sco are currently subjected to a FUV radiation field of $10-300\ \mathrm{G}_0$. For these levels of external UV radiation the mass loss rate due to external photo-evaporation at radii of $\sim100$ AU will be $\sim$$10^{-9}-10^{-8}\ \mathrm{M}_{\odot}\mathrm{yr}^{-1}$, which is of the same order of magnitude as the accretion rate (see, e.g. \citealt{Facchini2016,Haworth2017,Haworth2018}).
Given the age of the region, stars with even earlier spectral types might have been present but are now evolved, as shown by the red supergiant Antares, implying that in the past the region was subject to a more intense UV flux than it is at the present.

\emph{Magnetic disk winds versus viscous evolution:}
In Section \ref{sec: comparing to observations} we have shown that the observations in Lupus match with our viscously spreading disk models with $\alp = 10^{-3}-10^{-4}$.  
However, we cannot exclude an alternative interpretation in which the observed \rgas\ and stellar ages of individual disks could be reproduced by models in which disk evolution is driven by disk winds with a suitable choice of parameters. 
Figure \ref{fig: rgas evolution w observations} should therefore not be considered a confirmation of viscous evolution in disks.
To properly distinguish between whether viscous stresses or disk winds are the dominant driver of disk evolution requires observing viscous spreading (or lack thereof), i.e., that the average disk grows in size over time. Additionally, a search for disk winds in the objects discussed here could allow us to quantitatively compare how much specific angular momentum is extracted by disk winds and how much is transported outward by viscous stresses.

\emph{Episodic accretion:}
In this work we have assumed a simple prescription of viscous evolution, where viscosity in the disk is described by a single parameter \alp, which is kept constant in time and does not vary with radius. 
In reality this is likely to be a too simplistic view. For example, there is an increasing amount of evidence that stellar accretion is episodic rather than the smooth process assumed in this work (see, e.g. \citealt{Audard2014} for a review). 
It is likely however that episodic accretion is most important in the early phases when the star is being assembled, and probably less in the later Class II phase (see, e.g. \citealt{Costigan2014,Venuti2015}). 

If accretion were also episodic in the Class II phase, the growth of the disk size is likely also to be episodic, rather than the smooth curves shown in this work. However, in order to reproduce the average observed accretion rate, the episodic accretion rate averaged over time should still match the smooth accretion rate assumed in this work. 
Observationally, we cannot perform an average over time since the variational timescales, if any exists in the class II phase, are longer than what can be practically measured; observational studies find that accretion is modulated on the rotational period of the star \citep{Costigan2014, Venuti2015} but no variation on longer timescales. However, averaging over a sample of similar sources is mathematically equivalent to the average over time since they will be at different stages of their duty cycle.
Overall, the values for \alp\ discussed in this work thus should be intended as some kind of average over its variations in space and time.

Related to the topic of episodic accretion is the connection, at an early age, between the disk and its surrounding envelope. Material is accreted from this envelope onto the disk, with current evidence indicating that this could still be ongoing into the Class I phase (see, e.g. \citealt{Yen2019}). While this might affect our results at early ages (0.1-0.5 Myr) it is unlikely to change our results at a later age when accretion from the envelope onto the disk has stopped and it would be equivalent to changing the initial disk mass or the initial disk size.

\section{Conclusions}
\label{sec: conclusions}

In this work we used the thermochemical code \texttt{DALI} to examine how the extent of the CO emission changes with time in a viscously expanding disk model and investigate how well this observed measure of the gas disk size can be used to trace viscous evolution.
Below, we summarize our conclusions:
\begin{itemize}
    \item Qualitatively the gas outer radius \rgas\ measured from the $^{12}$CO emission of our models matches the signatures of a viscously spreading disk: \rgas\ increases with time and will do so at a faster rate if the disk has a higher viscous \alp\ (i.e. when it evolves on a shorter viscous timescale).
    \item For disks with high viscosity ($\alp \geq 10^{-2}$), the combination of a rapidly expanding disk with a low initial disk mass ($\mdisk \leq 2\times10^{-4}\ \msun$) can result in the observed outer radius decreasing with time, due to CO being photo-dissociated in the outer disk. 
    \item For most of our models, \rgas\ is up to $\sim$$12~\times$ larger than the characteristic size \rc\ of the disk, with the difference being larger for more massive disks. As a result, measuring \alp\ directly from observed \rgas\ will overestimate the true \alp\ of the disk by up to an order of magnitude. 
    
    \item Current measurements of gas outer radii in Lupus can be explained using viscously expanding disks with $\alp = 10^{-4}-10^{3}$ that start out small ($\rinit = 10$ AU). The exceptions are IM Lup (Sz 82) and HK Lup (Sz 98), which require either a higher $\alp \approx 10^{-2}$ or a larger initial disk size of $\rinit = 30-50$ AU to explain their large gas disk size. 
    
    \item Chemical depletion of CO through grain-surface chemistry has only minimal impact on the \rgas\ if measured from $^{12}$CO emission, but can significantly reduce \rgas\ at 5-10 Myr if measured from more optically thin tracers like $^{13}$CO.
\end{itemize}

We have shown that measured gas outer radii can be used to trace viscous spreading of disks and that models that fully simulate the observations are an essential part in linking the measured gas outer radius to the underlying physical size of the disk. Our analysis shows that current observations in Lupus are consistent with most disks starting out small and evolving viscously with low \alp. However, most sources lie within an age range of 1-3 Myr, too narrow to confirm that disk evolution is only driven by viscosity. We can therefore not rule out that disk winds are contributing to the evolution of the disk. Future observations should focus on expanding the available sample of observe gas disk sizes to other star-forming regions both younger and older than Lupus, to conclusively show if disk are viscous spreading and confirm whether viscosity is the dominant physics driving disk evolution.

\begin{acknowledgements}
We thank the referee for a very careful reading of the manuscript.
LT and MRH are supported by NWO grant 614.001.352. GR acknowledges support from the Netherlands Organisation for Scientific Research (NWO, program number 016.Veni.192.233). ADB acknowledges support from NSF Grant\#1907653. Astrochemistry in Leiden is supported by the Netherlands Research School for Astronomy(NOVA). All figures were generated with the \texttt{PYTHON}-based package \texttt{MATPLOTLIB} \citep{Hunter2007}. This research made use of Astropy,\footnote{http://www.astropy.org} a community-developed core Python package for Astronomy \citep{astropy:2013, astropy:2018}.
\end{acknowledgements}

\bibliographystyle{aa}
\bibliography{references}

\begin{appendix}

\section{Disk mass evolution}
\label{app: disk mass evolution}

\begin{figure}
    \centering
    \includegraphics[width=\columnwidth]{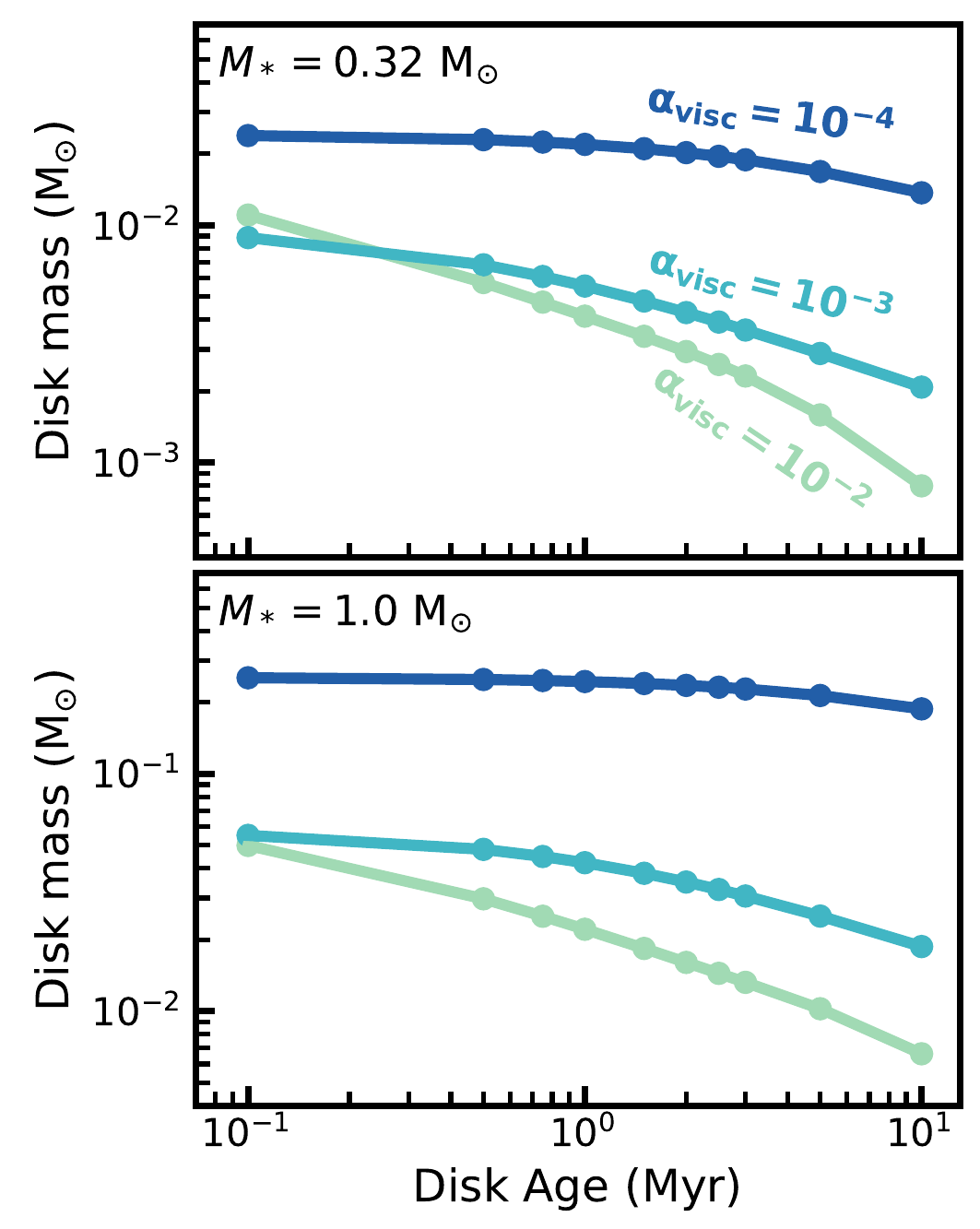}
    \caption{\label{fig: disk mass evolution} Evolution of the disk mass for the models with $\mstar = 0.32\ \msun$ and $1.0\ \msun$. Evolution of the disk mass for models with $\mstar = 0.1\ \msun$ is shown in Figure \ref{fig: disk parameter evolution}. Colors show models with different \alp. Note the order of magnitude difference between the $\mstar = 0.32\ \msun$ models (top panel) and $\mstar = 1.0\ \msun$ models (bottom panel). }
\end{figure}

\section{ $^{12}$CO radial intensity profiles}
\label{sec: 12CO radial profiles (all)}
\begin{figure*}[hptb]
    \centering
    {$^{12}$CO 2\,-\,1 intensity profiles -- non-depleted models}
    \includegraphics[width=\textwidth,clip,trim={1.3cm 1.1cm 2cm 1.6cm}]{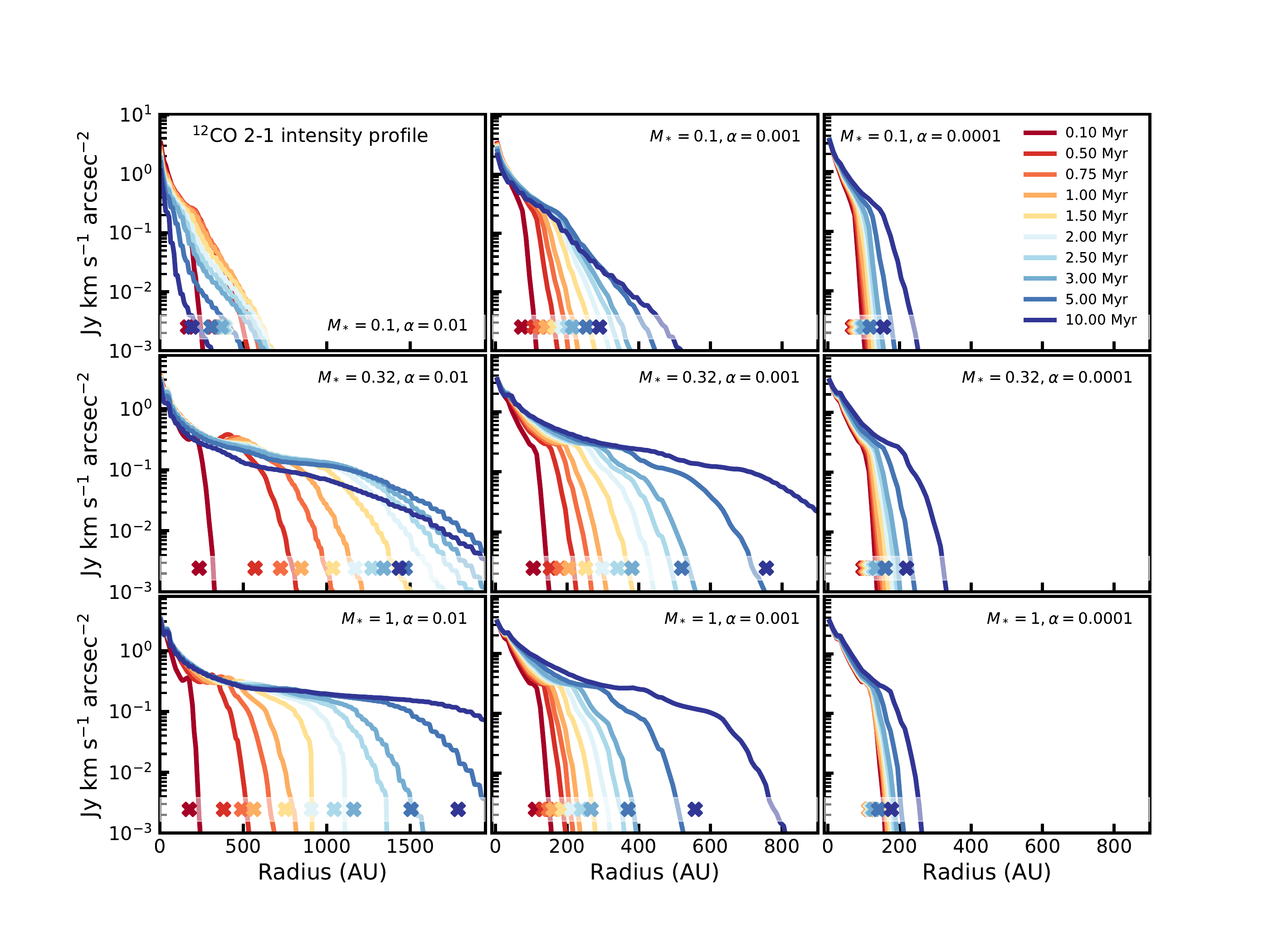}
    \caption{\label{fig: 12CO profiles non-depleted models} Time evolution of the $^{12}$CO intensity profiles for our grid of models. Rows show models with equal stellar mass, columns show models with equal $\alpha_{\rm visc}$. Colors, going from red to blue, show the different time steps. For each model, the radius enclosing 90\% of the total flux is marked by a cross. Note that a low stellar mass corresponds to a low stellar accretion rate, using the observational relation shown in Figure 6 in \cite{alcala2017}. Also note that, due to the setup, a low $\alpha_{\rm visc}$ corresponds to a high viscous time $(t_{\rm visc})$ and a high initial mass ($M_{\rm init}$).}
\end{figure*}

\section{Outer radii based $^{13}$CO emission}
\label{sec: 13CO results}
\begin{figure*}[hptb]
    \centering
    {$^{13}$CO 2\,-\,1 intensity profiles -- non-depleted models}
    \includegraphics[width=\textwidth,clip,trim={1.3cm 1.1cm 2cm 1.6cm}]{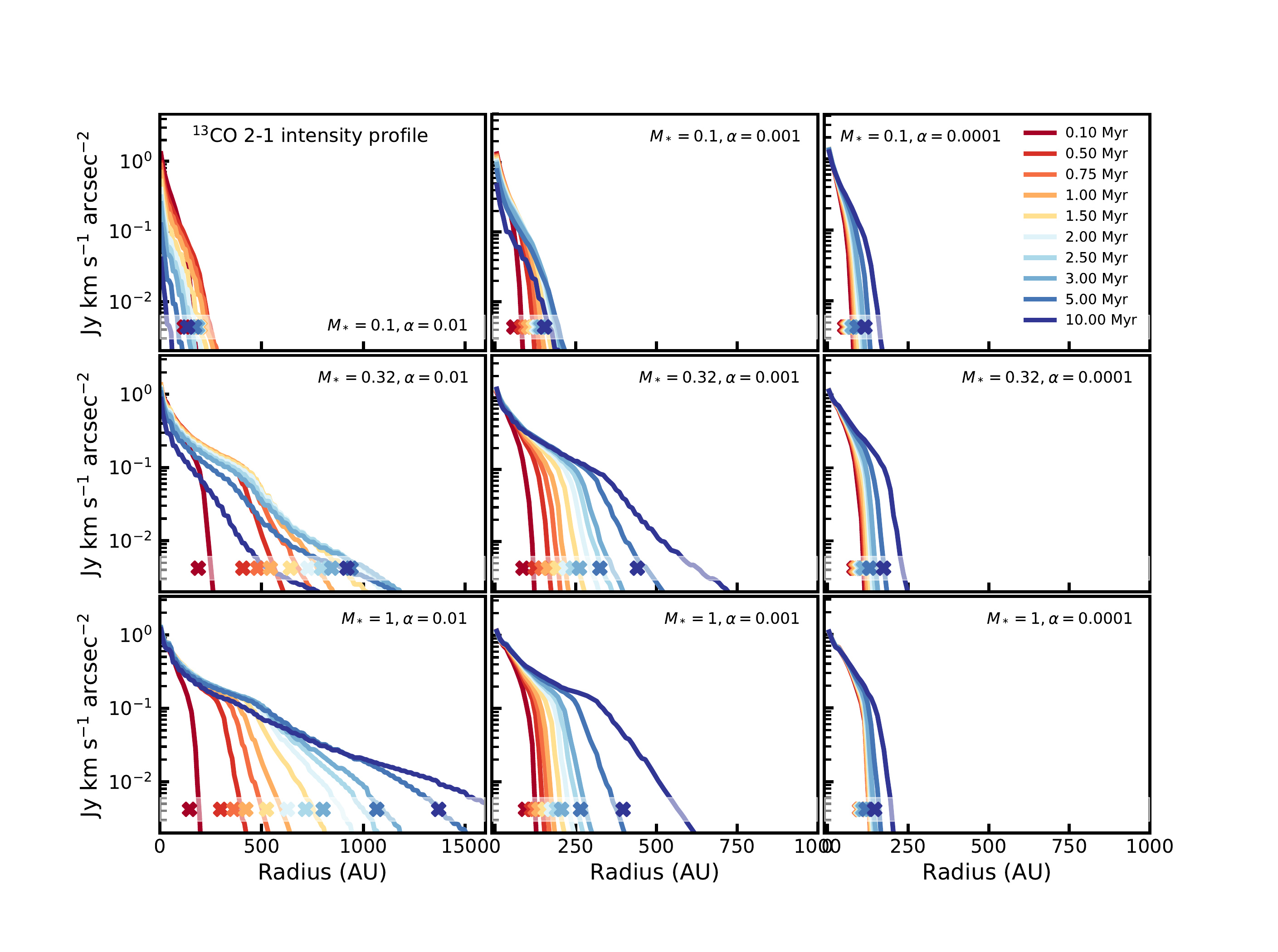}
    \caption{\label{fig: 13CO profiles non-depleted models} Time evolution of the $^{13}$CO intensity profiles for our grid of models. Rows show models with equal stellar mass, columns show models with equal $\alpha_{\rm visc}$. Colors, going from red to blue, show the different time steps. For each model, the radius enclosing 90\% of the total flux is marked by a cross. Note that a low stellar mass corresponds to a low stellar accretion rate, using the observational relation shown in Figure 6 in \cite{alcala2017}. Also note that, due to the setup, a low $\alpha_{\rm visc}$ corresponds to a high viscous time $(t_{\rm visc})$ and a high initial mass ($M_{\rm init}$).}
\end{figure*}

\begin{figure}[htb]
    \centering
    \includegraphics[width=\columnwidth]{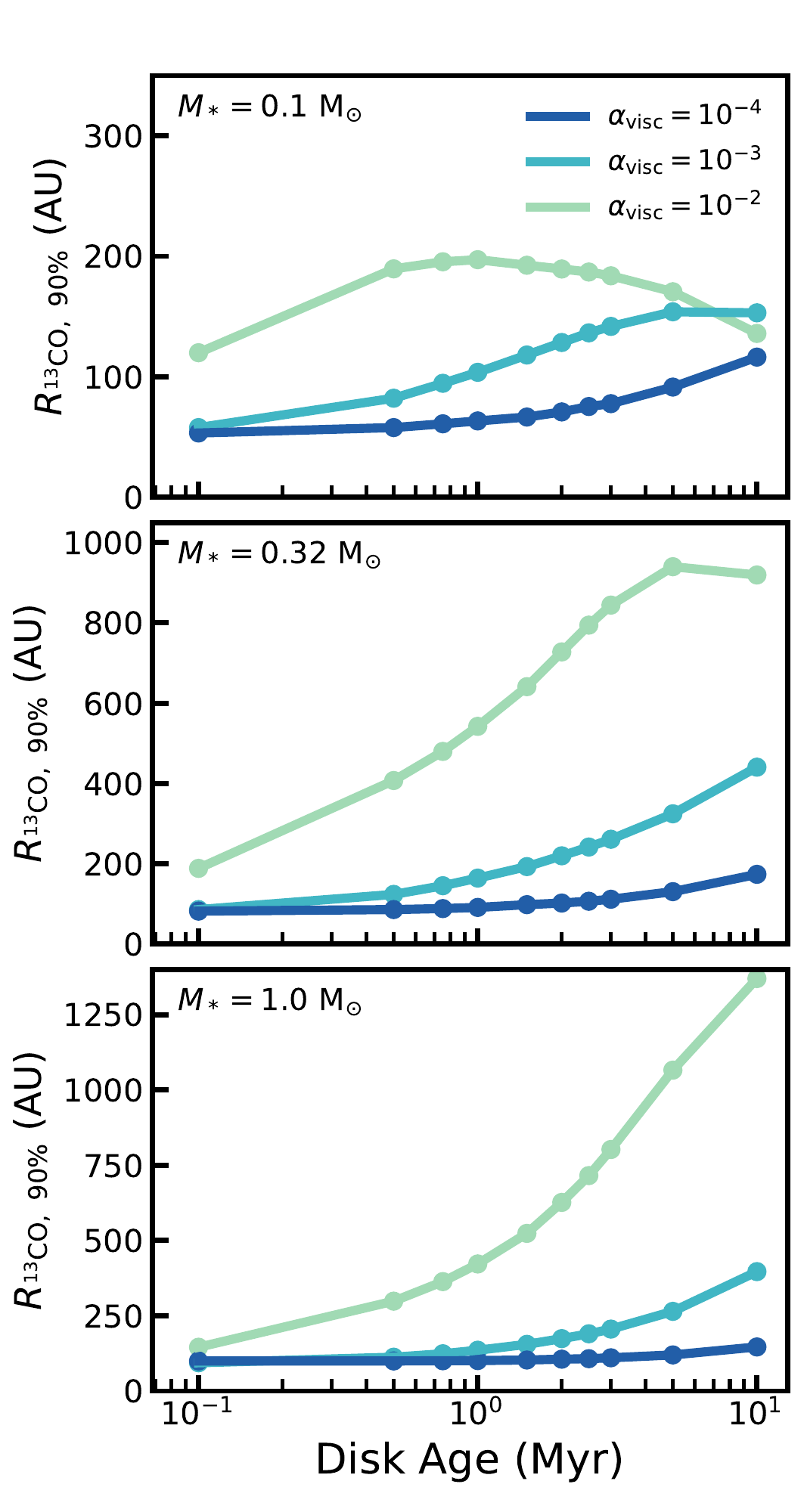}
\caption{\label{fig: 13CO outer radius evolution} Disk ages vs. gas outer radii (\rgas), measured from $^{13}$CO 2\,-\,1 emission.  Top, middle and bottom panels show models with different stellar masses. Colors indicate the \alp\ of the model. }
\end{figure}

\begin{figure}[htb]
    \centering
    \includegraphics[width=\columnwidth]{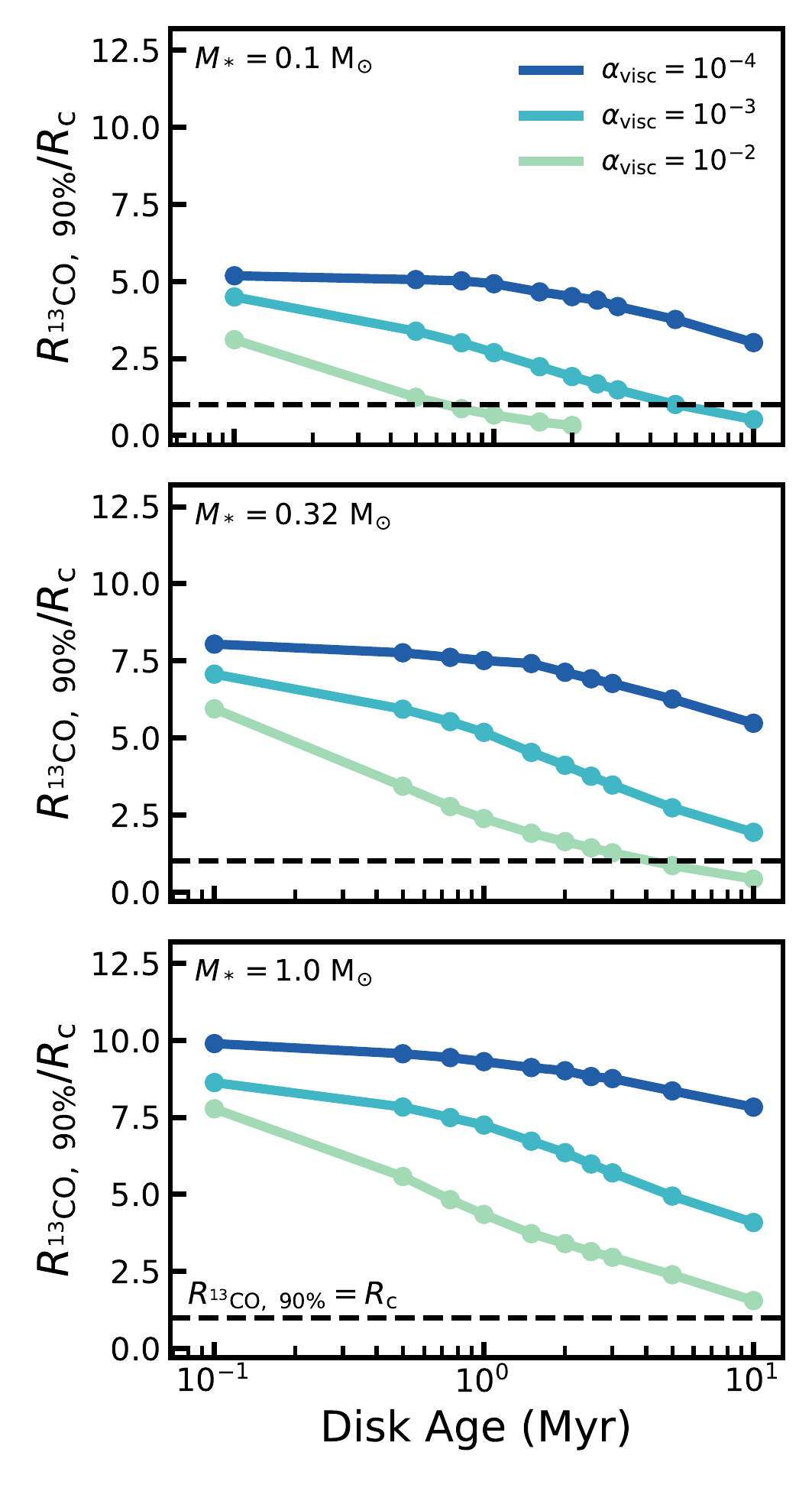}
\caption{\label{fig: 13CO outer radius traces rc} 
The ratio of gas outer radius ($R_{\rm ^{13}CO}$) over characteristic radius (\rc) vs. disk age. Gas outer radii are measured from $^{13}$CO 2\,-\,1 emission of our models. The black dashed line indicates where $R_{\rm ^{13}CO} = \rc$. Top, middle and bottom panels show models with different stellar masses. Colors represent models with different \alp.}
\end{figure}

\section{Observed sample}

\begin{table}[htb]
\centering
\caption{\label{tab: observations} Observations in Lupus and Upper Sco}
\def\arraystretch{1.5}
\begin{tabular*}{0.96\columnwidth}{l|cc|l}
\hline\hline
Name   &    $M_*$      &   age  &  \rgas \\
       & (M$_{\odot}$) &  (Myr) &   (AU) \\
\hline
\multicolumn{4}{c}{Lupus}\\
\hline
Sz 65  & $0.70 \pm 0.16$ & $1.3_{-0.8}^{+1.9}$ & $172 \pm 24$ \\
Sz 68  & $2.13 \pm 0.34$ & $0.8_{-0.4}^{+0.8}$   & $68  \pm 6$ \\
Sz 71  & $0.42 \pm 0.11$ & $2.0_{-1.2}^{+3.0}$   & $218 \pm 54$ \\
Sz 73  & $0.82 \pm 0.16$ & $4.0_{-2.4}^{+6.0}$   & $103 \pm 9$ \\
Sz 75  & $0.51 \pm 0.14$ & $0.8_{-0.5}^{+1.2}$   & $194 \pm 21$ \\
Sz 76  & $0.25 \pm 0.03$ & $2.0_{-1.0}^{+1.0}$   & $164 \pm 6$ \\
J15560210-3655282 & $0.46 \pm 0.12$ & $2.0_{-1.0}^{+1.0}$ & $110 \pm 3$ \\
IM Lup  & $1.10 \pm 0.0$ & $0.5_{-0.3}^{+0.8}$   & $388 \pm 84$ \\
Sz 84  & $0.18 \pm 0.03$ & $1.3_{-0.3}^{+5.1}$   & $146 \pm 18$ \\
Sz 129 & $0.80 \pm 0.15$ & $4.0_{-2.4}^{+6.0}$   & $140 \pm 12$ \\
RY Lup & $1.47 \pm 0.22$ & $2.5_{-1.3}^{+2.5}$   & $250 \pm 63$ \\
J16000236-4222145 & $0.24 \pm 0.03$ & $1.3_{-0.9}^{+1.3}$ & $266 \pm 45$ \\
MY Lup & $1.02 \pm 0.13$ & $10.0_{-5.0}^{+29.8}$ & $194 \pm 39$ \\
EX Lup & $0.56 \pm 0.14$ & $2.0_{-1.0}^{+1.0}$   & $178 \pm 12$ \\
Sz 133 & $0.63 \pm 0.05$ & $2.0_{-1.2}^{+3.0}$   & $238 \pm 66$ \\
Sz 91  & $0.47 \pm 0.12$ & $6.3_{-3.8}^{+9.5}$   & $450 \pm 80$ \\ 
Sz 98  & $0.74 \pm 0.20$ & $0.5_{-0.3}^{+0.8}$   & $358 \pm 52$ \\
Sz 100 & $0.18 \pm 0.03$ & $1.6_{-1.3}^{+0.9}$   & $178 \pm 12$ \\
J16083070-3828268 & $1.81 \pm 0.28$ & $2.5_{-1.3}^{+2.5}$ & $394 \pm 100$ \\
V1094 Sco & $0.47 \pm 0.14$ & $2.0_{-1.0}^{+1.0}$& $438 \pm 112$ \\
Sz 111  & $0.46 \pm 0.12$ & $5.0_{-3.0}^{+7.6}$  & $462 \pm 96$ \\
Sz 123A & $0.46 \pm 0.11$ & $2.0_{-1.0}^{+1.0}$  & $146 \pm 12$ \\
\hline
\multicolumn{4}{c}{Upper Sco}\\
\hline
J15534211-2049282 & $0.27_{-0.05}^{+0.07}$ & $(5-11)$ & $48.3_{-10.5}^{+9.6}$ \\
J16001844-2230114 & $0.19_{-0.04}^{+0.07}$ & $(5-11)$ & $58.2_{-13.4}^{+16.1}$ \\
J16020757-2257467 & $0.34_{-0.06}^{+0.07}$ & $(5-11)$ & $48.9_{-11.2}^{+14.1}$ \\
J16035767-2031055 & $1.05_{-0.11}^{+0.13}$ & $(5-11)$ & $26.3_{-7.0}^{+166.3}$ \\
J16035793-1942108 & $0.36_{-0.07}^{+0.07}$ & $(5-11)$ & $39.4_{-6.0}^{+7.7}$ \\
J16075796-2040087 & $0.45_{-0.09}^{+0.12}$ & $(5-11)$ & $31.3_{-2.6}^{+6.4}$ \\
J16082324-1930009 & $0.66_{-0.06}^{+0.08}$ & $(5-11)$ & $141.1_{-34.9}^{+30.8}$ \\
J16123916-1859284 & $0.51_{-0.08}^{+0.09}$ & $(5-11)$ & $154.2_{-24.7}^{+23.5}$ \\
J16142029-1906481 & $0.56_{-0.06}^{+0.05}$ & $(5-11)$ & $79.4_{-6.1}^{+6.3}$ \\
\hline
\end{tabular*}
\captionsetup{width=.92\columnwidth}
\caption*{\footnotesize{Stellar parameters derived from observations presented in \cite{alcala2014,alcala2017} (For the stellar age estimation, see \citealt{Andrews2018}}). Gas radii for Lupus from \cite{ansdell2018}. Stellar masses for Upper Sco taken from \cite{Barenfeld2016}. For the stellar age w use the 5-11 Myr stellar age of Upper Sco (see, e.g. \citealt{Preibisch2002,Pecaut2012}). For sources in Upper Sco \rgas\ were calculated from the fit to the observed $^{12}$CO intensity in \cite{barenfeld2017}.
}
\end{table}

\section{Local UV radiation field in Upper Sco}
\label{app: upper sco irradiation}

\begin{figure}
    \centering
    \includegraphics[width=\columnwidth,clip,trim={0.4cm 0.4cm 0.7cm 0.7cm }]{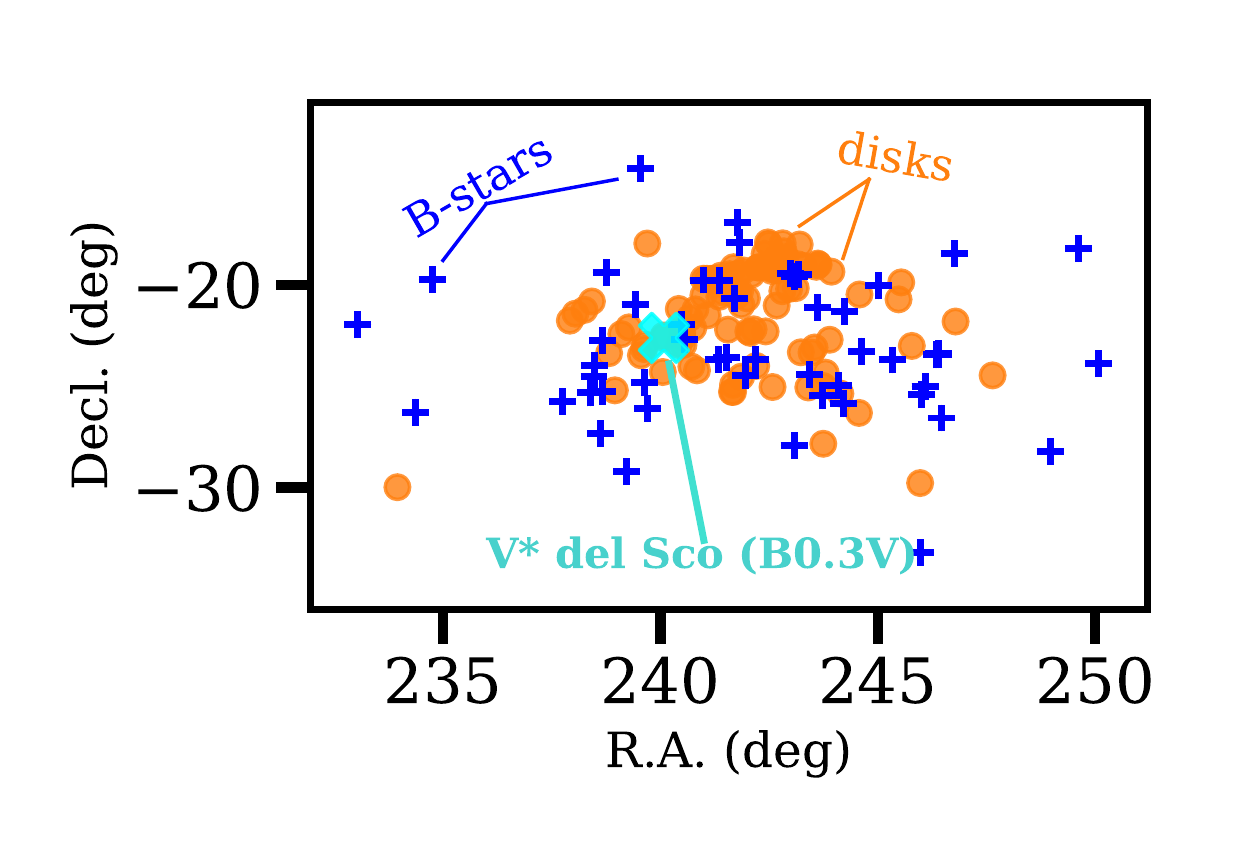}
    \caption{\label{fig: spatial distribution upper sco} Spatial distribution of the disks in Upper Sco (colored circles) and the 49 B stars (blue crosses). The lightblue cross denotes the location of V* del Sco, a B0.3V star that contributes significantly to the UV radiation in the region.
    }
\end{figure}

\begin{figure}
    \centering
    \includegraphics[width=\columnwidth,clip,trim={0.5cm 0.6cm 0.4cm 0.7cm}]{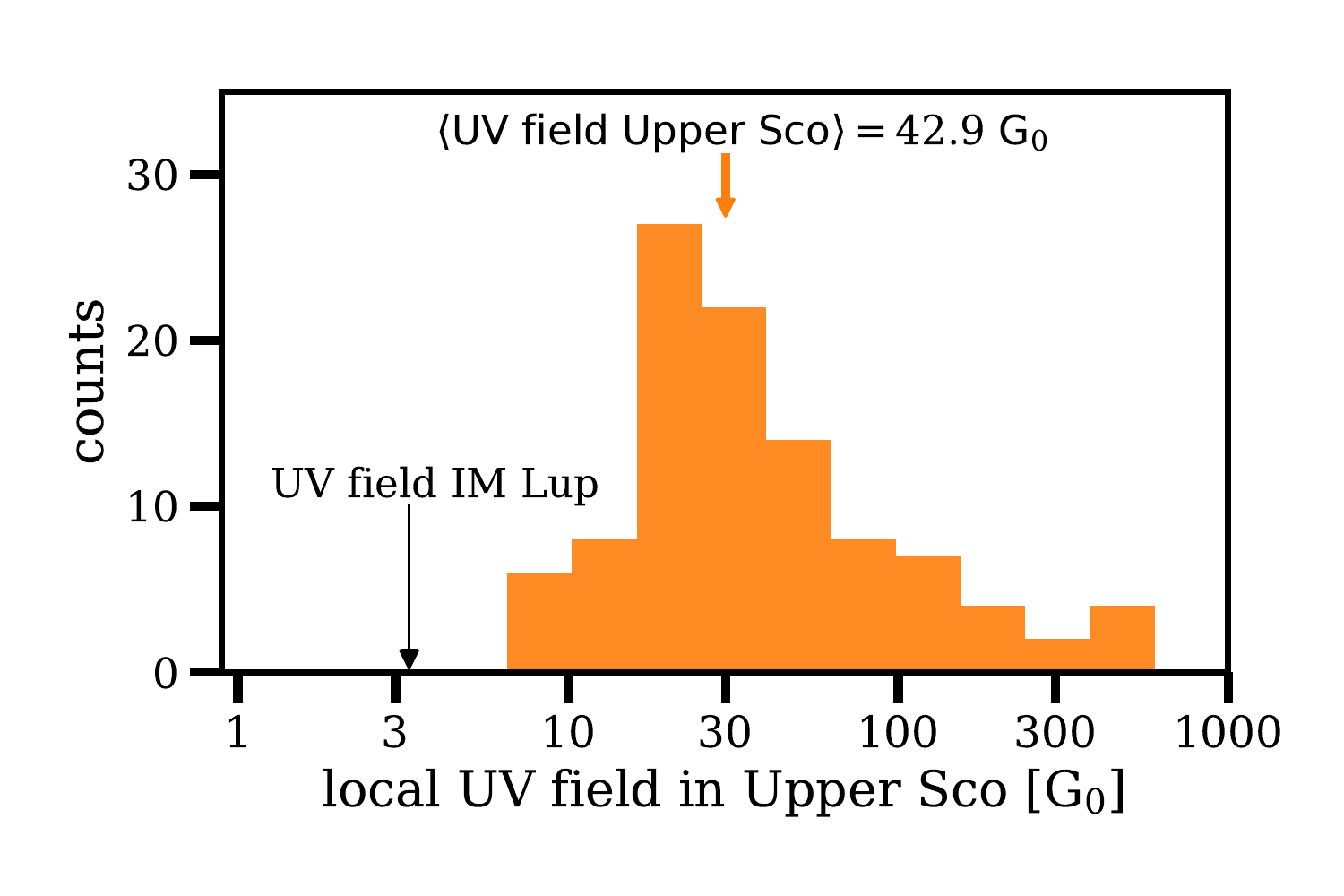}
    \caption{\label{fig: mdisk G0} Histogram of the local UV radiation field for disks in Upper Sco. Radiation fields were calculated for each disk using Equation \eqref{eq: UV field}. Orange arrow denotes the median UV radiation field. For comparison, the black arrow shows the external UV radiation field of IM Lup \citep{Cleeves2016}. }
\end{figure}

Being part of the nearest OB association, the disks in Upper Sco are likely to be subjected to high levels of irradiation. Here we quantify this these irradiation levels using the locations of known B stars \citep{deZeeuw1999} and the disks in Upper Sco (see Figure \ref{fig: spatial distribution upper sco}).

\cite{Barenfeld2016} presented ALMA observations of a sample of 106 stars in Upper Sco between spectral types of M5 and G2, selected based on the excess infrared emission observed by \textit{Spitzer} or \textit{WISE} \citep{Carpenter2006, LuhmanMamajek2012}. Parallaxes for 96 of these 106 stars were measured with \emph{Gaia} as part of its DR2 data release \citep{GaiaDR2_2018}. We use these parallaxes to calculate the distance to each of the stars. For the 10 stars where no parallax is available, we instead assume the distance to be 143 pc, which is the median distance to Upper Sco based on the DR2 data (see, e.g., \citealt{BailerJones2018,Wright2018,Damiani2019}).

In close proximity to these disk-hosting stars there are 49 stars with spectral type B1 and the region host one B0.3V star (V* del Sco) (see, e.g. \citealt{deZeeuw1999,deBruijne1999}). With the exception of the B0 star, these stars are also part of \emph{Gaia} DR2, allowing us to determine distances to these stars based on their parallaxes. The B0 star, V* del Sco, is too bright to be part of \emph{Gaia} DR2. For this star we use a distance of $224\pm24$ pc, which was calculated by \cite{Megier2009} based on interstellar Ca{\small\ II} lines.
Combining these positions and distances, we can now calculate, for each disk-hosting star, the relative distances between it and each of the B-stars. 

To calculate the local FUV radiation field, our approach the following:
We collect the spectral types of the 49 B stars from the catalog of \cite{deBruijne1999} and use them to compute their effective temperatures following \cite{HillenbrandWhite2004}. Based on these effective temperature we fit stellar masses using the stellar models from \cite{schaller1992}. Using the stellar masses we calculate the UV luminosity $L_{\rm UV,\ B\ star}$ produced by each star based on the relation presented in \cite{BuserKurucz1992}. Finally, for each disk in the sample from \cite{Barenfeld2016}, we calculate the local UV radiation $(F_{\rm UV,\ disk})$ by adding up the relative contributions of each of the nearby B stars:

\begin{equation}
\label{eq: UV field}
F_{\rm UV,\ disk} = \sum_{\rm B\ stars} \frac{L_{\rm UV,\ B\ star}}{|x_{\rm disk} - x_{\rm B\ star}|^2}
\end{equation}
Here $|x_{\rm disk} - x_{\rm B\ star}|$ denotes the relative distance between the disk and the B star. 

Figure \ref{fig: mdisk G0} shows a histogram of the local UV radiation $F_{\rm UV,\ disk}$ for the disks in Upper Sco. Levels range from $\sim 8$$\ \mathrm{G}_0$ to $\sim$$7\times10^{3}\ \mathrm{G}_0$, with a median $F_{\rm UV,\ disk} = 42.9\ \mathrm{G}_0$, confirming that these disks are subjected to high levels of external UV radiation. Even the lowest $F_{\rm UV,\ disk}$ ($\sim$8 G$_0$) is several times higher than the external UV field in Lupus ($\sim3\ \mathrm{G_0}$; \citealt{Cleeves2016}).

Due to the age of Upper Sco (5-11; see \citealt{Preibisch2002,Pecaut2012}), stars with spectral types earlier than B0.3, with lifetimes of a few up to 10 Myr might have been present in the region but are now evolved. It is therefore likely that external UV radiation in Upper Sco was higher in the past, suggesting that the $F_{\rm UV,\ disk}$ calculated here is a lower limit of what the disks in Upper Sco experienced during their lifetime.

\section{Implementing CO chemical depletion through grain-surface chemistry}
\label{app: implementation CO depletion}

Based on the results from \cite{Bosman2018b} an approximated description for CO grain surface chemistry has been developed. The description only traces the main carbon carriers, i.e., CO, CH$_3$OH, CO$_2$ and CH$_4$. Briefly, the approximation splits reactions into two groups, fast and slow reactions, and assumes that fast reactions can be calculated by equilibrium chemistry and that only the slow reactions need to be integrated explicitly. For a more detailed description, see the appendix of \krijtprep.
The approximation breaks down in the upper regions of the disk, where photodissociation of CO by UV photons becomes important. 
In these regions the chemistry included in \texttt{DALI} provides more accurate CO abundances.
We therefore split our disk models into two regions based on the outward CO column:
\begin{description}
    \item[$N_{\rm CO} < 10^{16}\ \mathrm{cm}^{-2}$:] 
    Here the outward CO column is too low for CO to self-shield against photo-dissociation. As a result the CO chemistry is accurately described by the photo-chemistry included in DALI and we therefore do not recompute the CO abundances.
    \item[$N_{\rm CO} \geq 10^{16}\ \mathrm{cm}^{-2}$:] Deeper in the disk CO is able to self-shield against photodissociation. Here grain-surface chemistry is able to convert CO into other species, thus lowering the gas-phase abundance of CO. For this region we recompute the CO abundances using the approximate grain-surface chemistry from \krijtprep.
\end{description}

\begin{figure*}[!thb]
    \centering
    \includegraphics[width=0.8\textwidth,clip,trim={1.1cm 0cm 0cm 0cm}]{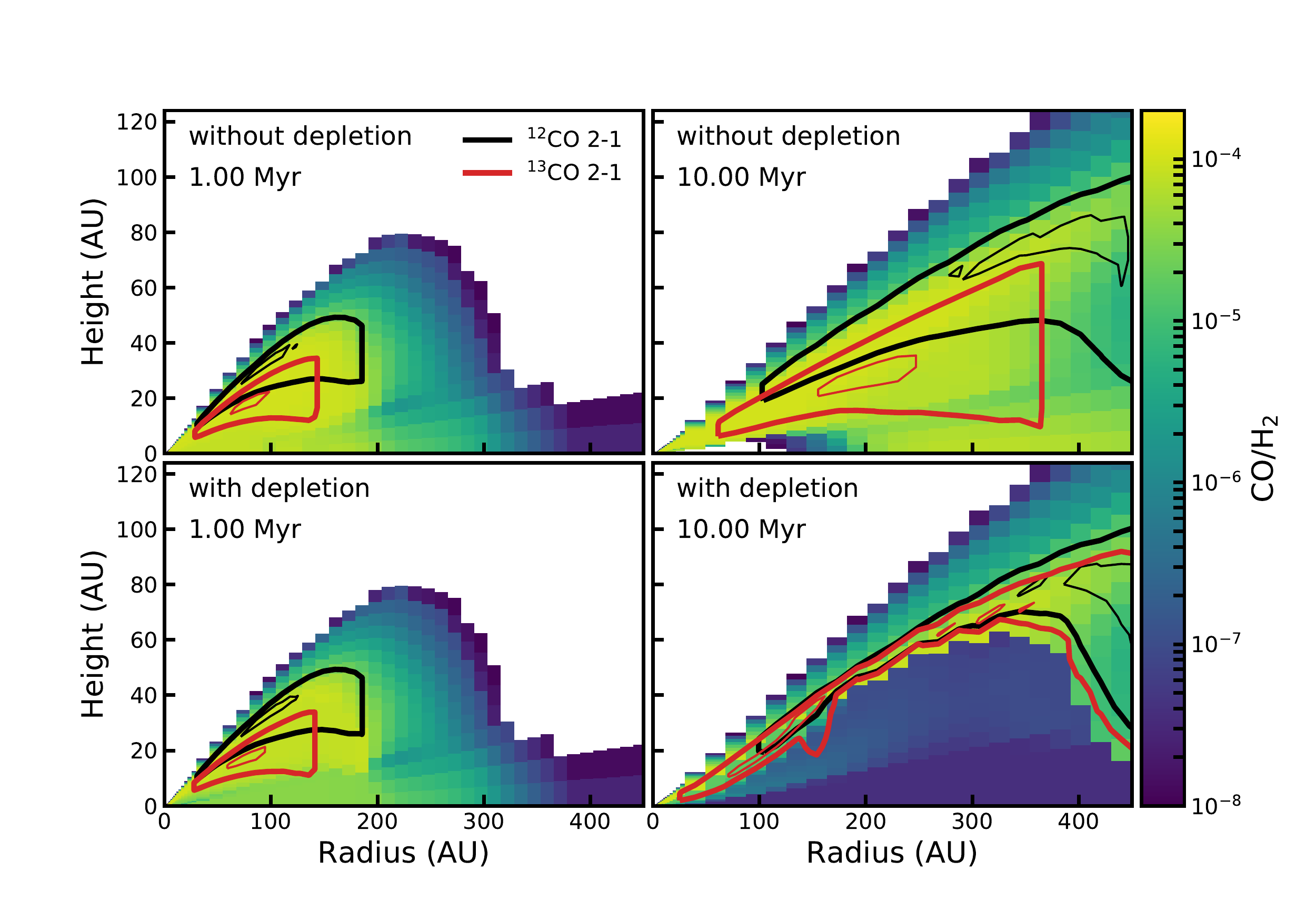}
    \caption{\label{fig: CO depletion in abundance} The effect of chemical depletion of CO through grain-surface chemistry on the CO abundance after 1 Myr (left panels) and 10 Myr (right panels). The example model shown here has $M_* = 0.32\ \mathrm{M}_{\odot}$ and $\alp = 10^{-3}$. Colors show the CO abundance with respect to H$_2$, where white indicates CO/H$_2 \leq 10^{-8}$. Black contours show the $^{12}$CO 2\,-\,1 emitting region, enclosing 25\% and 75\% of the total $^{12}$CO flux. Similarly, the red contours show the $^{13}$CO 2\,-\,1 emitting region. }
\end{figure*}

Figure \ref{fig: CO depletion in abundance} presents the CO abundance structure with and without including CO depletion through grain-surface chemistry. At 1 Myr, shown in left panels, the inclusion of grain-surface chemistry has little impact on the CO abundance. However, at 10 Myr CO has become significantly depleted (X[CO] $\leq 10^{-8}$) around the midplane of the disk.
We obtain the same conclusion as \cite{Bosman2018b}, namely that CO depletion only becomes significantly long timescales. Among our models we are only starting to see its effects after $\geq5$ Myr, indicating this process is most important for older star-forming regions like Upper Sco.

Figure \ref{fig: CO depletion in abundance} also shows the emitting layers of $^{12}$CO 2\,-\,1 and $^{13}$CO 2\,-\,1. The chemical conversion of CO into other species predominantly takes place around the midplane, while the CO emitting layer is higher up in the disk. For $^{12}$CO 2\,-\,1 the emitting layer lies predominantly in the region of the disk where the CO abundances are set by photo-dissociation and not grain-surface chemistry, and is therefore
only slightly affected by CO chemistry after 10 Myr. The emitting layer of $^{13}$CO 2\,-\,1 lies deeper in the disk and is therefore significantly affected by the conversion of CO.

\section{The effect of CO depletion on $^{13}$CO emission}
\begin{figure*}[hptb]
    \centering
    { $^{13}$CO 2\,-\,1 intensity profiles -- depleted models}
    \includegraphics[width=\textwidth,clip,trim={1.3cm 1.1cm 2cm 1.6cm}]{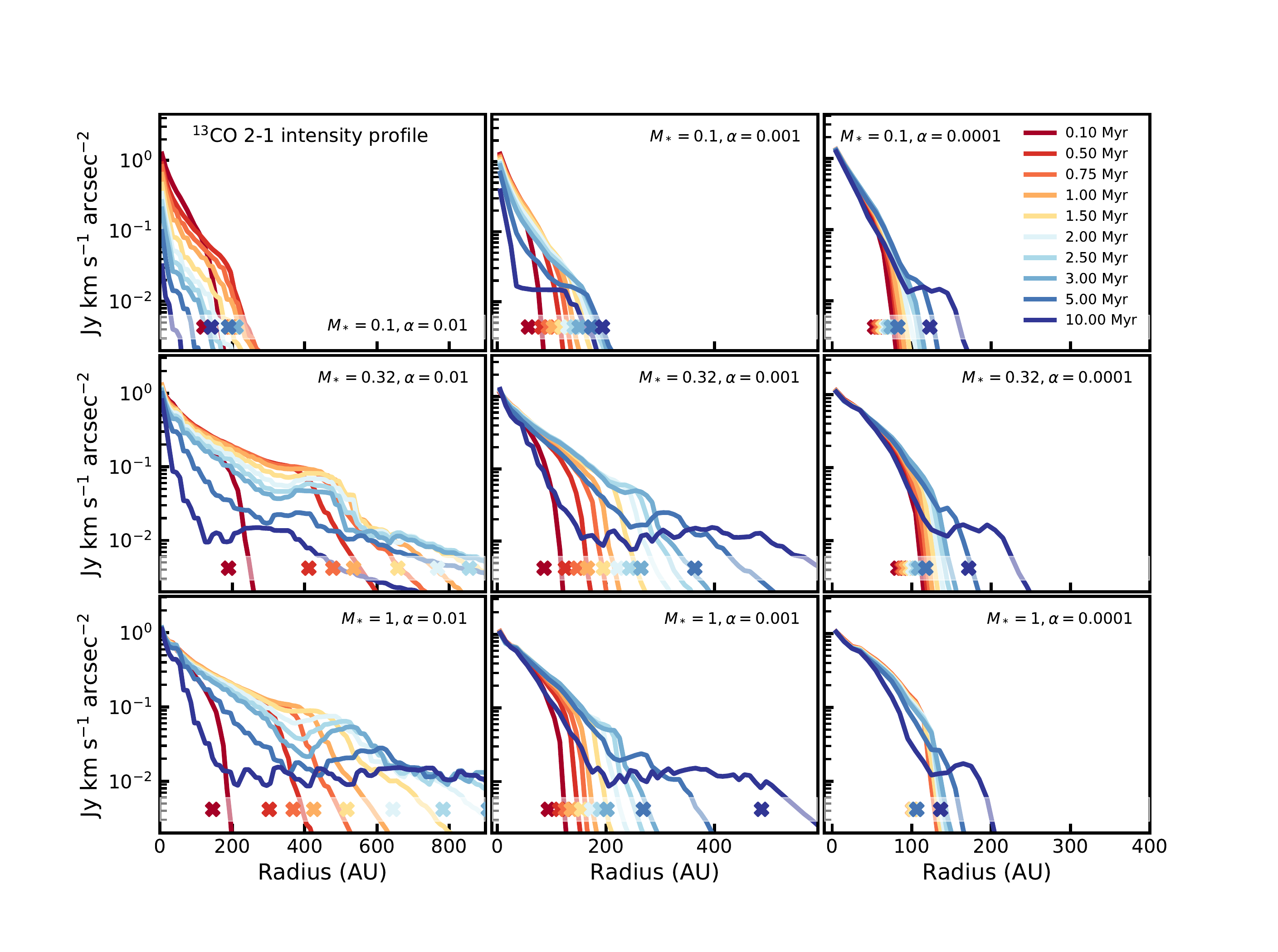}
    \caption{\label{fig: 13CO profiles depleted models} Time evolution of the $^{13}$CO intensity profiles for our grid of models, after including the effects of chemical depletion of CO through grain surface chemistry (see Sections \ref{sec: effect of CO depletion} and \ref{app: implementation CO depletion}). Rows show models with equal stellar mass, columns show models with equal $\alpha_{\rm visc}$. Colors, going from red to blue, show the different time steps. For each model, the radius enclosing 90\% of the total flux is marked by a cross. Note that a low stellar mass corresponds to a low stellar accretion rate, using the observational relation shown in Figure 6 in \cite{alcala2017}. Also note that, due to the setup, a low $\alpha_{\rm visc}$ corresponds to a high viscous time $(t_{\rm visc})$ and a high initial mass ($M_{\rm init}$).}
\end{figure*}

\end{appendix}
\end{document}